\PassOptionsToPackage{numbers, compress}{natbib}
\documentclass{article}

\usepackage{arxiv}
\paperkicker{Generalized Music Detection Framework}
\preprintnotice{arXiv Preprint}
\papercontact{\textbf{Corresponding:} \textcolor{RBBlue}{\texttt{xuyang@sustech.edu.cn}}; \textcolor{RBBlue}{\texttt{weili-fudan@fudan.edu.cn}}}

\usepackage[utf8]{inputenc} 
\usepackage[T1]{fontenc}    
\usepackage{hyperref}       
\usepackage{url}            
\usepackage{booktabs}       
\usepackage{amsfonts}       
\usepackage{nicefrac}       
\usepackage{microtype}      
\usepackage{xcolor}         

\usepackage{enumitem}
\usepackage{graphicx}
\usepackage{pgfplots}
\pgfplotsset{compat=1.18}
\usepackage{booktabs}
\usepackage{multirow}
\usepackage{amsmath}
\usepackage{booktabs}
\usepackage{array}    
\usepackage{url} 
\makeatletter
\g@addto@macro{\UrlBreaks}{\do\1\do\2\do\3\do\4\do\5\do\6\do\7\do\8\do\9\do\0\do\-\do\_}
\makeatother
\usepackage{ragged2e}

\usepackage{algorithm}
\usepackage{algpseudocode}
\usepackage{float}
\usepackage{multirow}
\usepackage{makecell}
\usepackage{amssymb}

\title{Beyond Artifacts: Towards Generalizable Synthetic Song Detection via Music-Intrinsic Features}

\author{%
{\large
\textbf{Yan Han}$^{1}$\RBEqual \quad
\textbf{Zhibin Wen}$^{2}$\RBEqual \quad
\textbf{Yuan Wang}$^{1}$ \quad
\textbf{Shuangrun Shao}$^{1}$ \quad
\textbf{Xiaobing Li}$^{1}$ \quad
\\[0.25em]
{\large
\textbf{Yang Xu}$^{2}$\RBCorresponding \quad
\textbf{Wei Li}$^{3}$\RBCorresponding}\\[0.55em]
{\fontsize{9.6}{11.3}\selectfont
$^1$Central Conservatory of Music \quad $^2$Southern University of Science and Technology \quad
$^3$Fudan University}}\\[0.45em]
{\small
\textbf{*}Equal contribution \quad
\S Corresponding authors} \\[0.55em]
\href{https://github.com/homura23/SOFIA}{%
  \raisebox{-0.18em}{\includegraphics[height=1.05em]{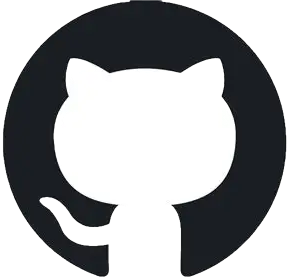}}%
  \hspace{0.25em}\textbf{Code: }%
  \texttt{https://github.com/homura23/SOFIA}%
}
\\[0.15em]
\href{https://huggingface.co/datasets/homura23/MUSIC8K}{%
  \raisebox{-0.18em}{\includegraphics[height=1.05em]{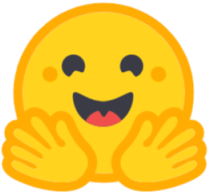}}%
  \hspace{0.25em}\textbf{Data: }%
  \texttt{https://huggingface.co/datasets/homura23/MUSIC8K}%
}
}

\begin{document}
\maketitle
\begin{abstract}

The rapid advancement of AI music generators highlights the urgent need for reliable Synthetic Song Detection (SSD).
Existing SSD methods often rely on low-level artifacts or fixed feature assumptions, struggling to capture generator-agnostic cues. To address this, we propose \textbf{Sofia} (\underline{\textbf{S}}ynthetic-song detecti\underline{\textbf{o}}n \underline{\textbf{f}}ramework via mus\underline{\textbf{i}}c fe\underline{\textbf{a}}tures), a flexible framework that models music-intrinsic attributes via feature-specific experts and an  adaptive Mixture-of-Experts (MoE) module. By configuring Sofia with representative Vocal, Audio-effect, Global structure features, and their combinations, we present their individual and complementary contributions. To comprehensively evaluate our framework, we further construct \textbf{MUSIC8K}, a challenging benchmark featuring lastest emerging generators and realistic audio perturbations. Experiments show that Sofia learns generator-agnostic representations from music-intrinsic features, improving the F1 score by 18.5 points over the strongest baseline on MUSIC8K-O while maintaining strong robustness.
\end{abstract}

\begin{figure}[t]
  \centering
  \includegraphics[width=0.45\columnwidth]{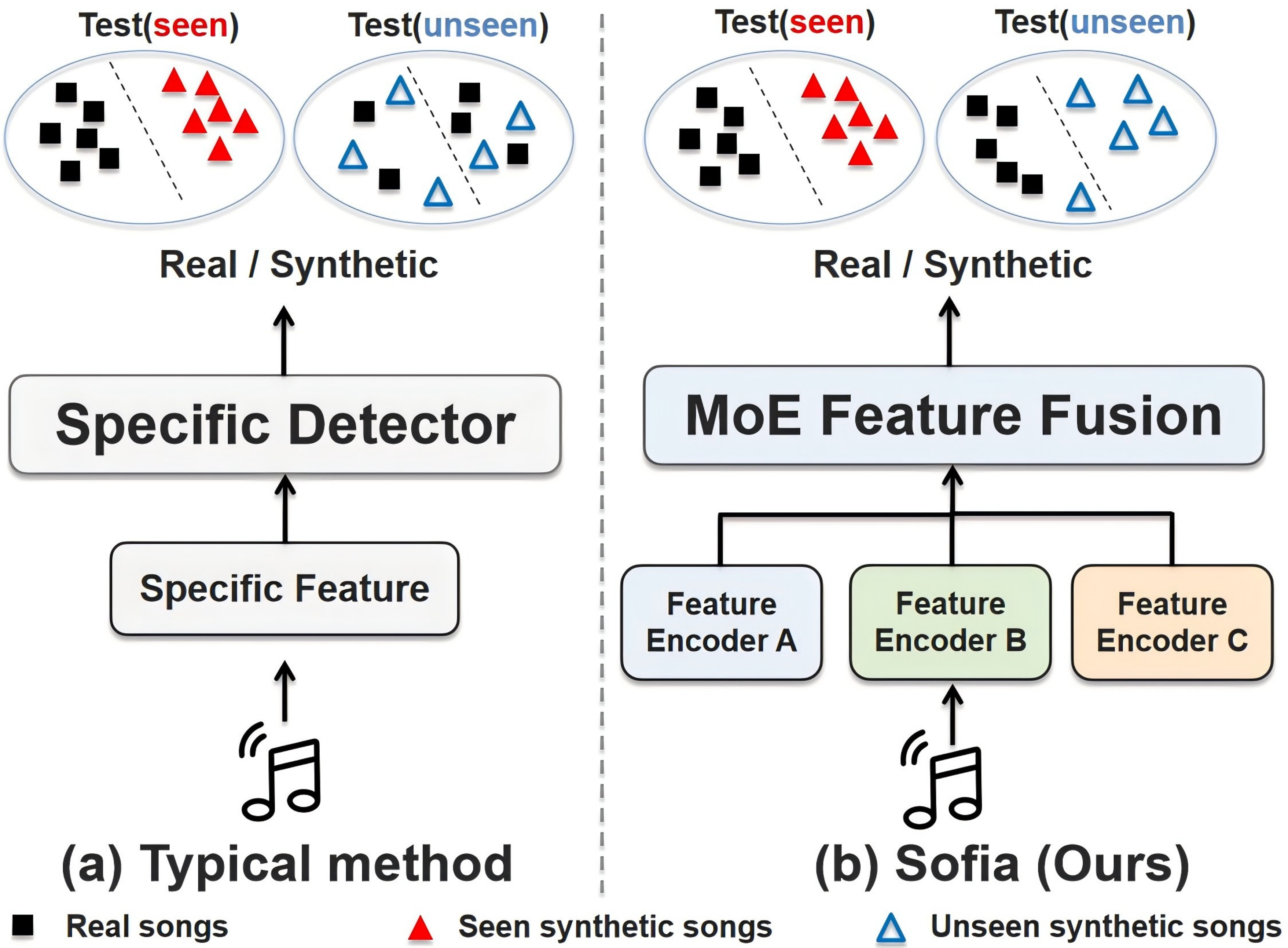}
  \caption{Comparison between typical detection methods and Sofia. (a) Typical methods rely on generator-specific artifacts. (b) Sofia is a flexible framework that models music-intrinsic features.}
  \label{fig:intro}
\end{figure}

\section{Introduction}
\label{sec:intro}

Recent advances in AI music generation have greatly improved the quality of synthetic songs~\cite{gong2026ace, yuan2025yue, ning2025diffrhythm}. Public platforms such as Suno\footnote{\label{fn:suno}\url{https://www.suno.ai/}} and Udio\footnote{\label{fn:udio}\url{https://www.udio.com/}} further lower the creation barrier, accelerating the production and dissemination of synthetic music. Recent blind listening tests show that 97\% of participants cannot distinguish fully synthetic music from human-composed tracks~\cite{Deezer2025AI}, suggesting that human perception alone is insufficient for reliable identification. As increasingly synthetic songs are uploaded to streaming platforms, they pose serious challenges to copyright protection, platform governance, artistic originality, and fair revenue distribution for human artists. These risks highlight the need for accurate, generalizable, and robust synthetic music detection methods.


Music is a multi-dimensional hierarchy with vocals, audio effects, and global musical structures~\cite{schedl2014music, kwiecien2024technical}. Although end-to-end Synthetic Song Detection (SSD)~\cite{zang2024ctrsvdd, zang2024singfake, xie2024fsd} is closely related to deepfake audio detection~\cite{yi2023audio}, as both aim to identify synthetic audio, existing deepfake audio detection methods are primarily designed for speech-oriented scenarios and therefore do not generalize well to music detection. Motivated by this gap, recent studies have developed methods specifically tailored to SSD. Spectrogram-based methods~\cite{rahman2024sonics, afchar2025fourier, pham2024deepfake} detect synthetic songs by modeling time-frequency artifacts, but they often rely on generator-specific cues and suffer significant performance degradation on unseen generators. Multimodal feature fusion approaches~\cite{frohmann2025double} combine lyrics with speech-model features, but they remains sensitive to the quality of lyrics transcriptions. CLAM~\cite{batramelody} is built on the hypothesis that synthetic music introduces subtle inconsistencies between vocal and instrumental elements, yet this assumption becomes increasingly fragile as music generators rapidly evolve. Overall, as illustrated in Figure~\ref{fig:intro}(a), despite recent progress, these synthetic music detection methods still face two critical challenges: achieving reliable generalization to the latest unseen music generators and maintaining robustness under audio perturbations.

These challenges also expose limitations in existing evaluation protocols. Although existing benchmarks such as SONICS~\cite{rahman2024sonics} and MoM~\cite{batramelody} provide large-scale and stylistically diverse synthetic music collections, the rapid evolution of synthetic music models leads to substantial data lag, making it difficult to evaluate detection methods on the latest generators.

To address these challenges, we propose \textbf{Sofia} 
(\underline{\textbf{S}}ynthetic-song detecti\underline{\textbf{o}}n 
\underline{\textbf{f}}ramework via mus\underline{\textbf{i}}c 
fe\underline{\textbf{a}}tures), a flexible framework for synthetic song detection based on music-intrinsic feature modeling. Sofia represents each song with a set of feature-specific experts, each capturing a distinct musical attribute, and adaptively combines these attributes through a Mixture-of-Experts (MoE) module~\cite{Jacobs1991AdaptiveMO}. Unlike methods built on a single artifact type or fixed feature assumption, Sofia can flexibly incorporate different numbers and types of music features, enabling systematic analysis of their individual and complementary contributions.

We instantiate Sofia with three representative music dimensions: \textbf{vocal features} (\textbf{V}), \textbf{audio-effect features} (\textbf{A}), and \textbf{global musical-structure features} (\textbf{G}). These features respectively characterize singing expression and timbre, spatial and mixing-related audio effects, and song-level musical organization such as rhythm, harmony, and section structure.

To evaluate synthetic song detection under latest generators, we construct \textbf{MUSIC8K}, a new benchmark that complements existing datasets with up-to-date generators and realistic perturbation scenarios. MUSIC8K consists of \textbf{MUSIC8K-O}, which contains songs generated by latest music generators for generalization evaluation, and \textbf{MUSIC8K-P}, which contains perturbed synthetic songs for robustness evaluation~\cite{sunday2025detecting}.

Finally, we conduct extensive experiments with different configurations under Sofia framework, including single-expert and multiple-experts combinations of V, A, G.
By combining music-intrinsic features, Sofia learns generator-agnostic representations and achieves state-of-the-art generalization to the latest generators and robustness against common audio manipulations. On MUSIC8K-O, Sofia improves the F1 score by \textbf{18.5} points over the strongest baseline. Our main contributions are as follows:

\begin{itemize}
    \item We propose \textbf{Sofia}, a flexible SSD framework that models music-intrinsic features with feature-specific experts and adaptively fuses them through a MoE module, enabling strong generalization to rapidly evolving AI music generators.
    
    \item We construct \textbf{MUSIC8K}, a new benchmark for evaluating SSD under latest music generators, covering both lastest generators and realistic audio perturbations.
    
    \item We conduct extensive experiments with different variants under the Sofia framework. Sofia achieves state-of-the-art generalization to the latest unseen generators, improving the F1 score by \textbf{18.5} points over the strongest baseline on MUSIC8K-O, while maintaining strong robustness.
\end{itemize}

    
    

\section{Related Work}
\label{sec:related_work}

\begin{figure*}[t]
  \centering
  \includegraphics[width=0.9\linewidth]{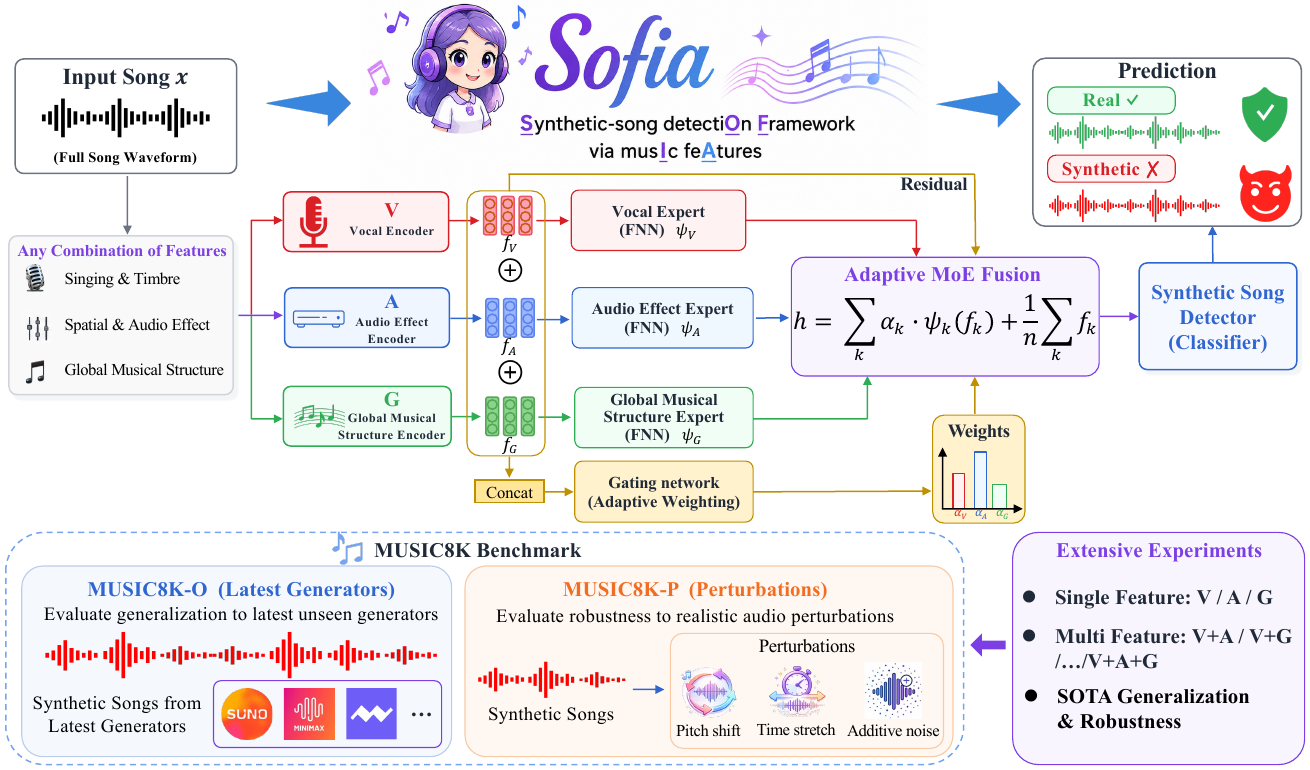}
  \caption{\textbf{Overview of our proposed Sofia framework.} Sofia extracts vocal, audio-effect, and global musical-structure features from input songs, processes them with specialized experts, and adaptively fuses their outputs through a Mixture-of-Experts module for real/synthetic classification. The framework supports flexible feature combinations and is evaluated on MUSIC8K under both latest unseen-generator and perturbation-based settings.
}
  \label{fig:main}
\end{figure*}

\paragraph{Detection Methods.} Early synthetic audio detection methods commonly convert waveforms into frequency-domain representations, such as LFCC~\cite{zhou2011linear}, CQT~\cite{brown1991calculation}, and Mel-spectrograms~\cite{stevens1937scale, allen2005unified}, and feed them into neural classifiers for binary prediction~\cite{zhang2021one,jung2022aasist,yi2023audio,pham2024deepfake}. More recent studies further improve detection by explicitly modeling temporal structures or leveraging pre-trained audio encoders. For example, SpecTTTra proposes the Spectro-Temporal Tokens Transformer to capture long-range temporal dependencies~\cite{rahman2024sonics}. Self-supervised audio representations have also been explored for music and vocal detection: Sharma ~\cite{sharma2025deepfake} use Whisper embeddings for vocal-based detection, while MoM-CLAM~\cite{batramelody} adopts a contrastive learning framework built on MERT~\cite{yizhi2023mert} and Wav2Vec2~\cite{baevski2020wav2vec} encoders.


Existing detectors often rely on low-level acoustic artifacts, making them prone to generator-specific overfitting and weak generalization to unseen generators. In contrast, Sofia models music-intrinsic cues across vocal, audio-effect, and global musical-structure dimensions, capturing structured interactions among vocals, instruments, effects, rhythm, and harmony for more robust and generalizable synthetic song detection~\cite{kwiecien2024technical}.

\paragraph{Datasets for Synthetic Song Detection.}

Several datasets have been proposed for synthetic music and singing voice detection. SingFake~\cite{zang2024singfake} provides paired real and synthetic vocal clips across multiple languages and singers, using real instrumental accompaniments to facilitate artifact analysis. CtrSVDD~\cite{zang2024ctrsvdd} further extends controlled singing voice deepfake detection with large-scale parameterized synthesis and metadata annotations. FSD~\cite{xie2024fsd} studies fake song detection in Chinese songs generated by multiple synthesis and conversion techniques. For full-song synthetic music detection, SONICS~\cite{rahman2024sonics} provides a large-scale benchmark for end-to-end SSD, containing over 97,000 full-length music tracks. It includes real-world audio collected from YouTube and synthetic songs generated by early music generation systems, such as Suno v2/v3/v3.5, Udio-32, and Udio-130. The more recent MoM benchmark~\cite{batramelody}, which contains 130,435 full-length songs from generators including Suno, Udio, Riffusion\footnote{\url{https://www.riffusion.com/}}, DiffRhythm~\cite{ning2025diffrhythm}, and Yue~\cite{yuan2025yue}.


Existing datasets are also limited: vocal-oriented datasets do not capture full-song structure, while full-song benchmarks rarely include the latest generators or realistic audio perturbations. MUSIC8K addresses these gaps with full-song evaluation for both the latest music generators generalization and robustness under realistic audio perturbations.





\section{Method}
\label{sec:method}



\subsection{Problem Setup}

Given a music sample $x$, synthetic music detection aims to classify whether $x$ is synthetic or real. Each sample is associated with a binary label $y \in \{0,1\}$, where $y=1$ denotes synthetic music and $y=0$ denotes real music. Formally, the task is defined as learning a detector $\mathcal{F}_{\Theta}$ that maps an input music sample $x$ to a binary prediction:
\begin{equation}
\hat{y}=\mathcal{F}_{\Theta}(x), \quad \hat{y}\in\{0,1\}.
\end{equation}

\subsection{Overview of Framework}

As shown in Figure~\ref{fig:main}, to improve generalization beyond generator-specific artifacts, we propose \textbf{Sofia}, a synthetic music detection framework that models complementary music-intrinsic cues (e.g., vocal features, audio-effect features, or global musical-structure features). Sofia consists of three main components: music-intrinsic feature extraction, a mixture-of-experts module, and a final prediction head.


\subsubsection{Music Feature Extraction}


Music synthesized by different generators often exhibits distinct feature biases across musical dimensions, such as vocals, audio effects, and global structure. Therefore, the music feature extraction module in Sofia is designed to be flexible, allowing different combinations of feature encoders to be incorporated. This design enables Sofia to construct an effective feature-encoding configuration for diverse generators and improves its generalization in SSD.


Formally, given a song sample $x$ with $n$ music features, we employ a set of encoders $\{E_k\}_{k=1}^{n}$ to extract feature-specific representations. To alleviate the inherent discrepancies across various encoders, a preprocessing stage is implemented to normalize and align the scale of each feature for a shared embedding space:
\begin{equation}
f_{k} = P_k\!\left( 
\frac{E_k(x)}{\|E_k(x)\|_2+\epsilon}
\right), \quad k=1,\ldots,n,
\label{eq:feature_extract}
\end{equation}
where $\epsilon$ is a small constant for numerical stability, and $P_k$ denotes a feature-specific projector that unifies the $f_{k}$ dimension across encoders. This preprocessing step mitigates discrepancies across encoders and yields a balanced input for the subsequent MoE module. 

\subsubsection{Music Feature Fusion}
Even songs from the same generator can exhibit substantial sample-level variations due to differences in genres, music styles. Therefore, static feature fusion strategies, such as direct concatenation or learning fixed global weights, can hardly capture discriminative cues that remain effective across diverse samples.

We assign an independent expert to each music feature and perform song-level aggregation through a mixture-of-experts (MoE) module~\cite{Jacobs1991AdaptiveMO}. The MoE module learns sample-dependent weights for different experts, enabling the detector to emphasize the most salient music features for each input. By adaptively integrating expert-specific representations, Sofia can better accommodate feature variations across songs and improve generalization to unseen generators.


Specifically, each expert is associated with one feature embedding $f_k$ and maps it into a feature-oriented discriminative representation through a feed-forward network (FFN) $\psi_k(\cdot)$. This design enhances feature-specific awareness and enables each expert to focus on the discriminative cues within its corresponding music feature. We implements $\psi_k(\cdot)$ as a single-layer network to mitigate overfitting by overly deep FFNs.

The integration of the expert-specific discriminative representation is applied by a gating network. It works as an expert weight routing, and is formally implemented as a multi-layer perceptron (MLP) $G_\theta(\cdot)$.
It evaluates the global content of each song from the original feature embeddings, and adaptively modulates the weights of each expert:
\begin{equation}
\begin{aligned}
\mathbf{z} &= \mathrm{Concat}\!\left(f_{1}, \ldots, f_{n}\right), \\
\alpha &= \mathrm{Softmax}\!\left(G_\theta(\mathbf{z})\right)
\end{aligned}
\label{eq:gating}
\end{equation}

where $z$ denotes the global feature embedding obtained via concatenation, and the weights satisfy $\sum_{k=1}^{n}\alpha_k=1$. Each $\alpha_k$ represents the instance-specific contribution of the corresponding expert, reflecting the feature-level composition of the input song. This input-dependent aggregation allows Sofia to adaptively emphasize salient discriminative cues across different tracks and generators.


To mitigate over-reliance on any single expert, we introduce a residual connection that preserves the original feature information and maintains contextual integrity. The fused representation $h$ used for prediction is computed as:
\begin{equation}
\mathbf{h} =
\sum_{k=1}^{n} \alpha_k \, \psi_k(f_k)
+
\frac{1}{n}\sum_{k=1}^{n} f_k .
\label{eq:fusion}
\end{equation}


By combining adaptive expert-specific representations with the residual feature information, $\mathbf{h}$ provides a unified discriminative representation for the subsequent classifier, improving robustness and generalization across diverse inputs.

\subsubsection{Prediction Head}
The fused embedding $\mathbf{h}$ is mapped by an output projection $O(\cdot)$, then $\ell_2$-normalized for numerical stability, and finally passed through a linear classifier followed by a softmax.
The predicted probability distribution $p \in \mathbb{R}^{2}$ over real and synthetic classes is computed as:
\begin{equation}
p = \mathrm{Softmax}\!\left(
W_c
\frac{O(\mathbf{h})}{\|O(\mathbf{h})\|_2+\epsilon}
+ b_c
\right)
\label{eq:prediction_head_full_simplified}
\end{equation}
where $W_c$ and $b_c$ denote the weight matrix and bias term of the linear classifier, respectively.
The predicted label is obtained from the class with the highest probability in $p$.


\subsection{Sofia Variants}

Sofia supports flexible instantiations by incorporating different combinations of music-intrinsic feature groups. In this work, we consider three complementary groups: Vocal (V), Audio-effect (A), and Global musical-structure features (G). This flexibility allows Sofia to construct detector variants that emphasize different musical dimensions, thereby better accommodating feature biases across music generators.

We instantiate Sofia with both single-feature and multi-feature configurations. The single-feature variants include Sofia-V1 (Wav2Vec2), Sofia-V1 (RawNet2), Sofia-A1 (FxPP), Sofia-G1 (MuQ), and Sofia-G1 (MERT). For feature fusion, we consider three full variants: Sofia-G2 (MuQ+MERT), Sofia-VAG (Concat) and Sofia-VAG (MoE), both of which integrate all three feature groups. Specifically, RawNet2~\cite{tak2021end} and Wav2Vec2~\cite{baevski2020wav2vec} model vocal-related cues, Fx-Encoder++~\cite{yeh2025fx} extracts audio-effect features, and MuQ~\cite{zhu2025muq} and MERT~\cite{yizhi2023mert} capture global musical-structure representations. Detailed variant configurations, music feature modeling, and implementation settings are provided in Appendices~\ref{sec:sofia_variants},~\ref{sec:music_feature_modeling}, and~\ref{sec:implementation_details}, respectively.

\subsection{Few-shot Ability}
\label{sec:few_shot_ability}

We further study whether Sofia can be adapted to a specific generator by few-shot learning. We denote this per-generator adaptation setting as \textbf{Sofia-VAG-FT}, which is initialized from the Sofia-VAG (MoE) base model and fine-tuned separately for each target generator. Given $N$ samples from a target generator, we jointly fine-tune the post-encoder components, including the branch projectors, the MoE-based fusion module, the output projection, and the final classification head. Formally, the trainable parameters are
\[
\Theta_{\mathrm{FT}} =
\{P_k\}_{k=1}^{n}
\cup \{\psi_k\}_{k=1}^{n}
\cup G_\theta
\cup O
\cup \{W_c,b_c\},
\]
where the encoders $\{E_k\}_{k=1}^{n}$ remain fixed.

\section{Experiments and Results}

\subsection{Datasets}
We construct the \textbf{MUSIC8K} dataset to evaluate generalization under recent generators and robustness under audio perturbations. It contains two subsets: \textbf{MUSIC8K-O} and \textbf{MUSIC8K-P}. MUSIC8K-O is used for generalization (cross-generator) evaluation. It contains 5{,}543 songs generated by recent commercial and open-source music generators, including ACE-Step 1.0, ACE-Step 1.5~\cite{gong2026ace}, Mureka O1, Mureka v9, Suno v5, Suno v5.5, Minimax 2.6, and HeartMuLa~\cite{yang2026heartmula}. The construction pipeline of MUSIC8K-O is provided in Appendix~\ref{sec:synthetic_generation_pipeline}. MUSIC8K-P is used for robustness evaluation. It is constructed from a subset of MUSIC8K-O and contains three splits manipulated through pitch shifting, time stretching, and additive noise. Each perturbation split contains 2{,}625 songs, and 7{,}875 perturbed songs in total. Details of the audio perturbation settings are provided in Appendix~\ref{sec:appendix_robustness}. We use the training split of SONICS~\cite{rahman2024sonics} for training, and evaluate all methods on SONICS, MoM~\cite{batramelody}, and MUSIC8K. Detailed source statistics are provided in Appendix~\ref{sec:Dataset Statistics}.

\begin{table*}[t]
\centering
\caption{Accuracy (\%) comparison of all baseline models across evaluation datasets. Best results are in bold.}
\label{tab:main_results}
\vspace{0.3em}
\renewcommand{\arraystretch}{1.05}
\setlength{\tabcolsep}{1.4pt}
\scriptsize
\resizebox{\textwidth}{!}{
\begin{tabular}{@{}l|*{8}{c}|*{5}{c}|*{5}{c}@{}}
\toprule
\textbf{Dataset}
& \multicolumn{8}{c|}{\textbf{MUSIC8K-O}}
& \multicolumn{5}{c|}{\textbf{MoM}}
& \multicolumn{5}{c}{\textbf{SONICS}} \\
\cmidrule{2-9} \cmidrule{10-14} \cmidrule{15-19}
\textbf{Metrics}
& \textbf{\makecell{Acestep\\1.0}}
& \textbf{\makecell{Acestep\\1.5}}
& \textbf{\makecell{Mureka\\O1}}
& \textbf{\makecell{Mureka\\V9}}
& \textbf{\makecell{Suno\\v5}}
& \textbf{\makecell{Suno\\v5.5}}
& \textbf{\makecell{Minimax\\2.6}}
& \textbf{Heartmula}
& \textbf{Riffusion}
& \textbf{DiffRhythm}
& \textbf{Yue}
& \textbf{\makecell{Suno\\v3}}
& \textbf{\makecell{Suno\\v4}}
& \textbf{\makecell{Suno\\v2}}
& \textbf{\makecell{Suno\\v3}}
& \textbf{\makecell{Suno\\v3.5}}
& \textbf{\makecell{Udio\\32}}
& \textbf{\makecell{Udio\\130}} \\
\midrule

RawNet2 (ICASSP 2021)
& 58.6 & 84.5 & 61.5 & 69.3 & 65.9 & 84.8 & 35.5 & 85.3
& 79.5 & 70.6 & 86.2 & 96.5 & 91.3
& 31.4 & 96.4 & 99.6 & 8.78 & 7.92 \\

ViT (ICLR 2021)
& -- & -- & -- & -- & -- & -- & -- & --
& -- & -- & -- & -- & --
& 82.0 & 99.0 & \textbf{100} & 53.0 & 99.0 \\

ConvNeXt (CVPR 2022)
& -- & -- & -- & -- & -- & -- & -- & --
& -- & -- & -- & -- & --
& 77.0 & 99.0 & 99.0 & 95.0 & \textbf{100} \\

EfficientViT (CVPR 2023)
& -- & -- & -- & -- & -- & -- & -- & --
& -- & -- & -- & -- & --
& 73.0 & 98.0 & \textbf{100} & 95.0 & \textbf{100} \\

SpecTTTra-$\gamma$ (ICLR 2025)
& 76.4 & 0.7 & 93.9 & 11.8 & 22.4 & 84.2 & 21.1 & 59.2
& 60.4 & 32.9 & 90.2 & 99.4 & 91.3
& 98.0 & 99.0 & \textbf{100} & 37.0 & \textbf{100} \\

SpecTTTra-$\beta$ (ICLR 2025)
& 46.3 & 0 & 77.5 & 0.1 & 1.8 & 39.4 & 4.6 & 12.6
& 53.4 & 26.7 & 89.2 & 99.2 & 56.5
& 87.0 & 99.0 & \textbf{100} & 62.0 & 99.0 \\

SpecTTTra-$\alpha$ (ICLR 2025)
& 81.8 & 2.2 & 93.8 & 12.4 & 22.9 & 88.1 & 34.5 & 50.6
& 62.8 & 37.8 & 91.4 & 99.2 & 82.6
& 78.0 & 99.0 & \textbf{100} & 96.0 & \textbf{100} \\

CLAM (TMLR 2025)
& 98.1 & 65.6 & 94.6 & 69.9 & 69.6 & 94.6 & 96.0 & 98.4
& 91.5 & 94.5 & 95.2 & 97.6 & 94.9
& 97.8 & 97.8 & 97.9 & 97.8 & 97.3 \\

\addlinespace
\midrule
\addlinespace

Sofia-A1(FxPP)
& 42.6 & 0 & 98.6 & 1.4 & 2.1 & 16.7 & 6.5 & 1.8
& 18.4 & 11.2 & 77.6 & 99.2 & 34.8
& 76.5 & 99.5 & 95.2 & 62.7 & 86.9 \\

Sofia-G1(MuQ)
& 99.4 & 86.9 & 95.6 & 78.7 & 64.7 & 96.9 & 97.7 & 97.8
& 95.3 & 98.7 & 98.4 & \textbf{100} & 95.6
& 78.7 & \textbf{100} & 96.5 & 98.2 & 52.1 \\

Sofia-G1(MERT)
& \textbf{100} & 43.9 & \textbf{99.5} & 71.6 & 31.8 & 96.9 & \textbf{99.3} & \textbf{99.4}
& \textbf{97.5} & 83.1 & 98.7 & 99.8 & \textbf{100}
& \textbf{100} & \textbf{100} & \textbf{100} & \textbf{100} & 98.7 \\

Sofia-V1(Wav2Vec2)
& 98.6 & 6.7 & 64.5 & 90.6 & 11.8 & 35.5 & 59.5 & 96.6
& 79.8 & 67.4 & 91.2 & 90.6 & 91.3
& 75.3 & 97.3 & 94.7 & \textbf{100} & 90.7 \\

Sofia-V1(RawNet2)
& 79.5 & 79.8 & 78.9 & 71.6 & 69.4 & 74.6 & 77.9 & 72.1
& 68.3 & 54.2 & 73.9 & 97.2 & 73.9
& 40.7 & 90.7 & 97.3 & 60.1 & 40.6 \\

Sofia-G2(MuQ+MERT)
& 98.7 & 55.4 & 94.3 & 49.4 & 11.7 & 81.5 & 92.4 & 97.9
& 87.5 & 72.5 & 97.5 & \textbf{100} & \textbf{100}
& 80.4 & \textbf{100} & 98.2 & 98.7 & 84.7 \\

Sofia-VAG (concat)
& \textbf{100} & 41.5 & 95.3 & 60.7 & 29.1 & 87.9 & 94.4 & 98.4
& 92.5 & 90.1 & 98.8 & \textbf{100} & 95.7
& 93.3 & \textbf{100} & \textbf{100} & \textbf{100} & 92.3 \\

\textbf{Sofia-VAG (MoE)}
& \textbf{100} & \textbf{91.2} & 94.8 & \textbf{85.5} & \textbf{83.2} & \textbf{97.3} & 95.4 & 98.7
& 92.3 & \textbf{98.8} & \textbf{98.9} & 99.5 & 99.6
& 97.4 & 99.8 & \textbf{100} & 99.8 & 98.8 \\

\bottomrule
\end{tabular}
}
\vspace{-0.6em}
\end{table*}

\begin{figure*}
  \centering
  \includegraphics[width=0.95\textwidth]{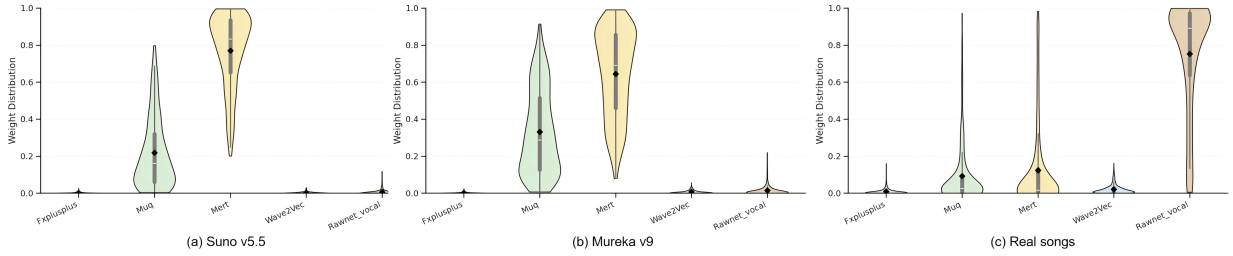}
  \caption{Dataset-level expert weights for (a) Suno v5.5, (b) Mureka v9, and (c) real songs. The black diamond denotes the mean expert weight, and the thick box inside each violin represents the interquartile range.
}
  \label{fig:dataset_weight}
\end{figure*}

\subsection{Baseline Methods}

We compare Sofia with representative baselines, including ConvNeXt~\cite{Liu2022ACF}, ViT~\cite{Dosovitskiy2020AnII}, EfficientViT~\cite{Liu2023EfficientViTME}, SpecTTTra-$\gamma$, SpecTTTra-$\beta$, SpecTTTra-$\alpha$~\cite{rahman2024sonics}, RawNet2~\cite{tak2021end}, and CLAM~\cite{batramelody}. Details of baselines are provided in Appendix~\ref{sec:baseline_details}.

\subsection{Results}

We adopt \textbf{F1-score} and \textbf{Accuracy} as evaluation metrics across all experiments.

\subsubsection{Evaluation on Existing Benchmarks}
Table~\ref{tab:main_results} reports per-generator accuracy, revealing that different Sofia variants exhibit selective superiority depending on their encoder configurations.
Specifically, Sofia-G1(MERT) performs strongly on most SONICS subsets and also obtains the best accuracy on Riffusion and Suno v4 in MoM. Meanwhile, Sofia-VAG (MoE) achieves the best accuracy on several challenging generators, including DiffRhythm and Yue in MoM, as well as Suno v3.5, Udio 32, and Udio 130 in SONICS. This supports that Sofia as a flexible framework for constructing detectors with different feature configurations.

Table~\ref{tab:f1_results} summarizes the overall F1-scores. On SONICS, CLAM obtains the highest F1-score of \textbf{99.3\%}, while Sofia-VAG (MoE) remains competitive with \textbf{97.4\%}. However, on MoM, Sofia-VAG (MoE) achieves the best F1-score of \textbf{98.5\%}, outperforming CLAM by \textbf{6.0 percentage points}. This indicates that multi-feature modeling in Sofia-VAG (MoE) is more reliable for more recent generators.





\begin{figure*}[t]
  \centering
  \includegraphics[width=0.85\linewidth]{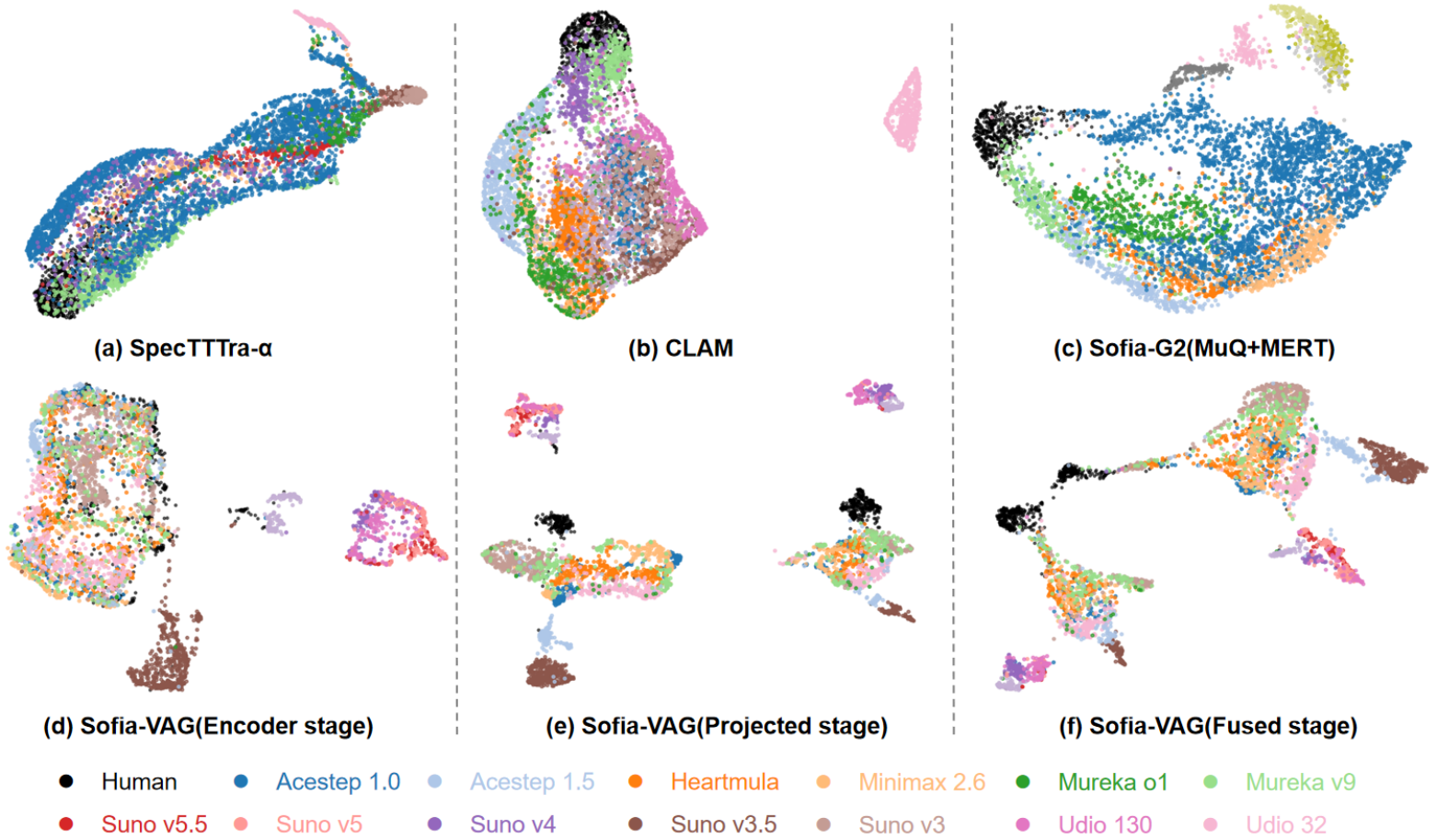}
  \caption{UMAP visualization~\cite{McInnes2018UMAPUM} of feature representations. (a)--(c) show the feature distributions of SpecTTTra-$\alpha$, CLAM, and Sofia-G2(MuQ+MERT). (d)--(f) show the feature distributions of Sofia-VAG at three architectural stages.}
  \label{fig:umap}
\end{figure*}

\subsubsection{Generalization to Latest Generators}
\begin{minipage}{0.43\linewidth}
MUSIC8K-O poses a substantially greater generalization challenge than existing benchmarks. 
As shown in Table~\ref{tab:f1_results}, all baselines drop from above \textbf{85\%} F1-score to below \textbf{70\%}; even the strongest baseline, CLAM, drops by \textbf{15.0} points. 
In contrast, all Sofia variants except Sofia-A1 match or outperform CLAM, with Sofia-VAG achieving the highest F1-score of \textbf{97.2\%}. 
More than half of the Sofia variants also maintain F1-scores around or above \textbf{90\%}. These results demonstrate Sofia's ability to generalize from legacy training data to emerging generators, highlighting its efficacy in capturing generator-agnostic cues.

Table~\ref{tab:main_results} shows that different Sofia variants perform best on different MUSIC8K-O generators. 
Compared with the strongest external baseline, Sofia variants improve accuracy by \textbf{1.9}, \textbf{6.7}, \textbf{4.9}, \textbf{20.7}, \textbf{13.6}, \textbf{2.7}, \textbf{3.3}, and \textbf{1.0} points on ACE-Step 1.0, ACE-Step 1.5, Mureka O1, Mureka v9, Suno v5, Suno v5.5, MiniMax 2.6, and HeartMuLa, respectively. 
Since these improvements come from different Sofia variants, the results suggest that different music-intrinsic features capture generator-specific artifacts.

\end{minipage}
\hfill 
\begin{minipage}{0.55\linewidth} 
    \raggedleft 
    \small
    \captionsetup{type=table,hypcap=false}
    \caption{F1-score comparison (\%) of existing baselines and different Sofia variants.}
    \label{tab:f1_results}
    \begin{tabular}{lccc}
    \toprule
    \textbf{Dataset}
    & \textbf{MUSIC8K-O}
    & \textbf{MoM}
    & \textbf{SONICS} \\
    \cmidrule(lr){2-4}
    \textbf{Metrics}
    & \multicolumn{3}{c}{\textit{F1 Score (\%)}} \\
    \midrule
    RawNet2 (ICASSP 2021) & 59.3 & 63.5 & 61.5 \\
    ViT (ICLR 2021) & -- & -- & 89.0 \\
    ConvNeXt (CVPR 2022) & -- & -- & 96.0 \\
    EfficientViT (CVPR 2023) & -- & -- & 95.0 \\
    SpecTTTra-$\gamma$ (ICLR 2025) & 63.2 & 80.4 & 88.0 \\
    SpecTTTra-$\beta$ (ICLR 2025)& 36.5 & 82.1 & 92.0 \\
    SpecTTTra-$\alpha$ (ICLR 2025) & 64.6 & 86.9 & 97.0 \\
    CLAM (TMLR 2025) & 78.7 & 92.5 & \textbf{99.3} \\
    
    \addlinespace
    \midrule
    \addlinespace
    
    Sofia-A1(FxPP) & 28.2 & 55.6 & 52.8 \\
    Sofia-G1(MuQ) & 95.3 & 98.4 & 88.4 \\
    Sofia-G1(MERT) & 89.2 & 96.7 & 93.9 \\
    Sofia-V1(Wav2Vec2) & 77.6 & 88.6 & 83.2 \\
    Sofia-V1(RawNet2) & 82.5 & 78.4 & 56.5 \\
    Sofia-G2(MuQ+MERT) & 87.5 & 93.2 & 95.1 \\
    Sofia-VAG (concat) & 88.1 & 94.5 & 97.2 \\
    \textbf{Sofia-VAG (MoE)} & \textbf{97.2} & \textbf{98.5} & 97.4 \\
    
    \bottomrule
    \end{tabular}
    \end{minipage}
Finally, the cross-variant comparison in Table~\ref{tab:f1_results} highlights the complementarity of music-intrinsic features. 
Although specialized variants can excel on specific generators---for example, Sofia-G1 (MERT) reaches nearly \textbf{100\%} accuracy on MiniMax 2.6 and HeartMuLa---Sofia-VAG achieves the best overall performance across datasets. 
Moreover, Sofia-VAG (MoE) consistently outperforms Sofia-VAG (concat), confirming that the gains come from adaptive MoE-based fusion rather than simply adding more features.

\subsubsection{Robustness under Audio Perturbations}
\begin{minipage}{0.38\linewidth}
As shown in Table~\ref{tab:audio_perturbations}, three audio perturbations are evaluated on MUSIC8K-P. Compared with its unperturbed MUSIC8K-O performance of \textbf{95.4\%} accuracy and \textbf{97.2\%} F1-score, Sofia-VAG (MoE) only drops by \textbf{1.3} accuracy points and \textbf{1.3} F1 points under time stretching. Under additive noise, it drops by \textbf{9.6} accuracy points and \textbf{7.8} F1 points, but still maintains the best F1-score of \textbf{89.4\%}. Pitch shifting affects Sofia-VAG(MoE) more noticeably, while Sofia-G2(MuQ+MERT) remains strong under this perturbation, suggesting that global musical-structure features are less sensitive to pitch changes. By integrating complementary features, Sofia-VAG reduces over-reliance on single feature to maintains stronger robustness under realistic audio perturbations.
\end{minipage}
\hfill 
\begin{minipage}{0.6\linewidth} 
    \raggedleft 
    \small
    \captionsetup{type=table,hypcap=false}
    \caption{Robustness comparison on MUSIC8K-P.}
    \label{tab:audio_perturbations}
    \begin{tabular}{lcccccc}
    \toprule
    \textbf{Dataset}
    & \multicolumn{2}{c}{\textbf{Pitch}} 
    & \multicolumn{2}{c}{\textbf{Stretch}} 
    & \multicolumn{2}{c}{\textbf{Noise}} \\
    \cmidrule(lr){2-3} \cmidrule(lr){4-5} \cmidrule(lr){6-7}
    \textbf{Metrics}
    & \textbf{Acc.} & \textbf{F1}
    & \textbf{Acc.} & \textbf{F1}
    & \textbf{Acc.} & \textbf{F1} \\
    \midrule
    
    RawNet2 (ICASSP 2021)
    & 80.4 & 72.6 & 80.3 & 72.6 & 72.4 & 66.9 \\
    SpecTTTra-$\gamma$ (ICLR 2025)
    & 54.9 & 25.7 & 56.5 & 29.4 & 56.9 & 34.5 \\
    SpecTTTra-$\beta$ (ICLR 2025)
    & 50.9 & 7.1 & 51.2 & 8.9 & 51.8 & 12.2 \\
    SpecTTTra-$\alpha$ (ICLR 2025)
    & 62.9 & 49.1 & 57.8 & 34.9 & 58.5 & 39.2 \\
    CLAM(TMLR 2025)
    & 80.6 & 88.4 & 81.2 & 88.9 & 61.5 & 69.9 \\
    
    \addlinespace
    \midrule
    \addlinespace
    
    Sofia-A1(FxPP) 
    & 32.8 & 19.3 & 32.9 & 19.5 & 29.9 & 17.8 \\
    Sofia-G1(MuQ) 
    & 83.1 & 86.9 & 83.1 & 87.1 & \textbf{85.8} & 89.3 \\
    Sofia-G1(MERT) 
    & 79.7 & 86.4 & 79.5 & 86.3 & 76.2 & 81.4 \\
    Sofia-V1(Wav2Vec2) 
    & 67.5 & 76.9 & 67.9 & 77.4 & 45.3 & 47.1 \\
    Sofia-V1(RawNet2) 
    & 68.4 & 78.4 & 67.9 & 77.4 & 56.7 & 66.2 \\
    Sofia-G2(MuQ+MERT) 
    & \textbf{89.2} & 92.3 & 89.2 & 92.3 & 72.4 & 76.6 \\
    Sofia-VAG (concat) 
    & 88.2 & \textbf{91.4} & 88.1 & 91.3 & 64.3 & 67.5 \\
    \textbf{Sofia-VAG (MoE)}
    & 83.1 & 87.1 & \textbf{94.1} & \textbf{95.9} & \textbf{85.8} & \textbf{89.4} \\
    
    \bottomrule
    \end{tabular}
\end{minipage}

\subsubsection{Analysis of Expert Weights}
\paragraph{Dataset-level}
Figure~\ref{fig:dataset_weight} shows the distribution of expert weights across Suno v5.5, Mureka v9, and real songs. For Suno v5.5 and Mureka v9, MERT and MuQ receive dominant weights. In contrast, RawNet2 receives higher weights in real songs, indicating that Sofia-VAG learns distinct weighting patterns for different sources. Other generators are provided in Appendix~\ref{sec:additional_expert_weights}.

\paragraph{Sample-level}

\begin{figure}
  \centering
  \includegraphics[width=0.7\columnwidth]{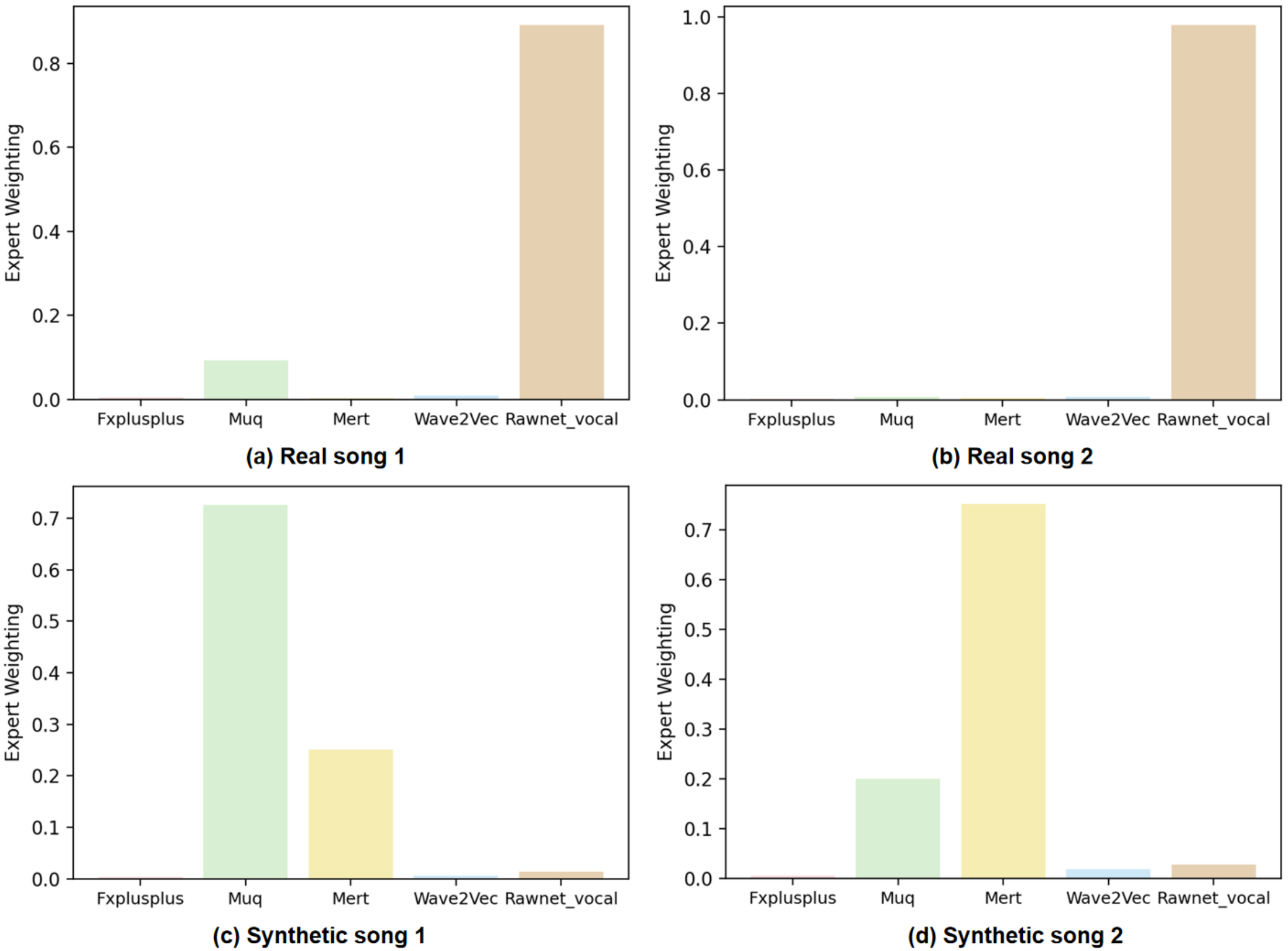}
  \caption{Sample-level expert weights for real and synthetic songs. (a)--(b) show two real-song examples. (c)--(d) show two Suno v5.5 synthetic-song examples.}
  \label{fig:single_weight}
\end{figure}

Figure~\ref{fig:single_weight} shows sample-level variations. Different samples activate different experts, the overall patterns remain consistent with the dataset-level distributions in Figure~\ref{fig:dataset_weight}: real songs assign higher weights to RawNet2, while synthetic songs rely more on MuQ and MERT. This indicates that Sofia-VAG captures both sample-specific differences and source-level feature tendencies.

\subsubsection{Analysis of Sofia Variants}

We further analyze different variants of \textbf{Sofia} to examine the effects of feature configurations and fusion strategies. For feature-level analysis, we keep the MoE-based fusion module fixed and remove one encoder at a time, yielding Sofia-VG (w/o A), Sofia-VAG (w/o MuQ), Sofia-VAG (w/o MERT), Sofia-VAG (w/o Wav2Vec2), and Sofia-VAG (w/o RawNet2). For fusion analysis, we keep the VAG features unchanged and replace MoE fusion with concatenation or simple weighting, where fixed global weights are learned for feature branches.

Table~\ref{tab:feature_configuration} shows that each feature contributes to Sofia-VAG on MUSIC8K-O. 
Sofia-VAG with MoE fusion achieves the best performance, reaching \textbf{95.4\%} accuracy and \textbf{97.2\%} F1-score, which suggests that vocal, audio-effect, and global musical-structure features offer complementary cues for detecting recent out-of-distribution generators. 
On MoM, Sofia-VAG (w/o RawNet2) performs best, with \textbf{98.9\%} accuracy and \textbf{99.5\%} F1-score, indicating that the optimal feature configuration can be generator-dependent. 
This highlights Sofia's flexibility in adapting feature combinations to different detection scenarios. The fusion comparison confirms the advantage of MoE fusion. 
On MUSIC8K-O, Sofia-VAG (MoE) improves over Sofia-VAG (simple-weight) by \textbf{8.7} accuracy points and \textbf{5.7} F1 points. This demonstrates that sample-dependent expert weighting better integrates heterogeneous music features and improves generalization over fixed fusion strategies.

\begin{table}
\centering
\caption{Performance comparison of different Sofia variants.}
\label{tab:feature_configuration}
\vspace{0.5em}
\renewcommand{\arraystretch}{0.85}
\setlength{\tabcolsep}{2.0pt}
\tiny
\resizebox{0.6\linewidth}{!}{
\begin{tabular}{lcccccc}
\toprule
\textbf{Dataset} 
& \multicolumn{2}{c}{\textbf{MUSIC8K-O}} 
& \multicolumn{2}{c}{\textbf{MoM}} 
& \multicolumn{2}{c}{\textbf{SONICS}} \\
\cmidrule(lr){2-3} \cmidrule(lr){4-5} \cmidrule(lr){6-7}
\textbf{Metrics}
& \textbf{Acc.} & \textbf{F1}
& \textbf{Acc.} & \textbf{F1}
& \textbf{Acc.} & \textbf{F1} \\
\midrule

Sofia-VG (w/o A)              & 82.5 & 88.5 & 95.5 & 97.5 & 97.4 & 96.9 \\
Sofia-VAG (w/o MuQ)           & 90.1 & 93.9 & 96.1 & 97.9 & 94.2 & 93.6 \\
Sofia-VAG (w/o MERT)          & 91.2 & 94.6 & 93.3 & 96.3 & 92.9 & 91.1 \\
Sofia-VAG (w/o Wav2Vec2)      & 84.9 & 90.3 & 95.0 & 97.3 & 95.5 & 94.7 \\
Sofia-VAG (w/o RawNet2)       & 95.1 & 97.1 & \textbf{98.9} & \textbf{99.5} & 94.6 & 94.1 \\
Sofia-VAG (concat)            & 81.9 & 88.1 & 94.5 & 96.9 & \textbf{98.1} & 97.2 \\
Sofia-VAG (simple-weight)     & 86.7 & 91.5 & 95.8 & 97.7 & 97.7 & 97.2 \\

\addlinespace
\midrule
\addlinespace

\textbf{Sofia-VAG (MoE)}
& \textbf{95.4} & \textbf{97.2}
& 97.2 & 98.5
& 97.8 & \textbf{97.4} \\

\bottomrule
\end{tabular}
}
\vspace{-0.6em}
\end{table}

\subsubsection{Representation Visualization}
Figure~\ref{fig:umap} provides feature-space comparisons between baseline methods and Sofia variants. First, the representations of SpecTTTra-$\alpha$ and CLAM are largely mixed across different sources, indicates that artifact-based or specific-assumption detectors are insufficient for separating latest unseen generators. Second, Sofia-G2(MuQ+MERT) and Sofia-VAG both show stronger separability than the baseline methods. However, Sofia-VAG further produces a more discriminative feature space. This comparison indicates that complementary multi-feature modeling provides more expressive representations. Third, the three stages of Sofia-VAG show a progressive improvement in feature-space separability. At the encoder stage, directly combining raw encoder features still leaves many sources mixed. After projection, the feature space becomes more compact and better structured, indicating that mapping heterogeneous encoder outputs into a unified embedding space is necessary. Finally, after MoE-based fusion, the separation between different generator groups, becomes clearer. This demonstrates that the adaptive fusion module further enhances discriminative feature organization and supports Sofia’s cross-generator detection ability.

\begin{table}[H]
\centering
\caption{Per-generator few-shot adaptation results on six latest MUSIC8K-O generators. Sofia-VAG-FT is initialized from Sofia-VAG (MoE), freezes all audio encoders, and fine-tunes only the fusion module and classification head with $N=50$ target samples.}
\label{tab:fewshot_target_results}
\vspace{0.2em}
\renewcommand{\arraystretch}{0.95}
\setlength{\tabcolsep}{3.5pt}
\small
\begin{tabular}{lccc}
\toprule
\textbf{Target Generator}
& \textbf{Sofia-VAG (MoE)}
& \textbf{Sofia-VAG-FT}
& $\Delta$ \textbf{F1} \\
\midrule
ACE-Step 1.5 & 90.23 & 99.73 & +9.51 \\
HeartMuLa    & 97.10 & 99.73 & +2.63 \\
MiniMax 2.6  & 90.23 & 98.26 & +8.03 \\
Mureka V9    & 95.13 & 99.35 & +4.22 \\
Suno v5.5    & 90.88 & 98.77 & +7.89 \\
Suno v5      & 80.21 & 98.64 & +18.43 \\
\bottomrule
\end{tabular}
\vspace{-0.8em}
\end{table}

\subsubsection{Per-generator Few-shot Adaptation}
\label{sec:per_generator_few_shot}
We further evaluate the few-shot ability of Sofia on recent generators from MUSIC8K-O. Specifically, we select six latest target generators, including ACE-Step 1.5, HeartMuLa, MiniMax 2.6, Mureka V9, Suno v5.5, and Suno v5. For each target generator, we initialize from the same Sofia-VAG (MoE) base model and train an independent Sofia-VAG-FT using only $N$ target samples, while keeping all audio encoders frozen.

Table~\ref{tab:fewshot_target_results} reports the per-generator adaptation results at $N=50$. Sofia-VAG-FT consistently improves the target-domain performance for all six generators, showing that few-shot adaptation can effectively strengthen Sofia for a specific generator. The improvement is especially pronounced on the weaker Suno v5 domain: the base Sofia-VAG (MoE) achieves \textbf{80.21\%} F1-score, whereas Sofia-VAG-FT increases it to \textbf{98.64\%}, yielding a gain of \textbf{18.43} points. Similar gains are also observed on ACE-Step 1.5, MiniMax 2.6, and Suno v5.5, where the target F1-score improves by \textbf{9.51}, \textbf{8.03}, and \textbf{7.89} points, respectively.

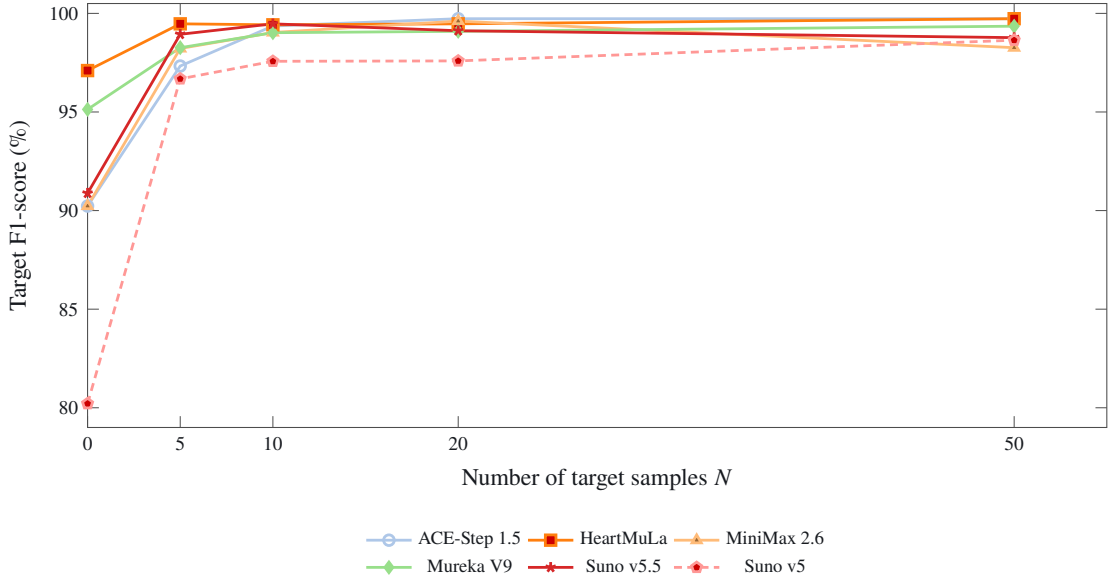
\begin{figure}[H]
\centering
\vspace{-0.3em}
\begin{tikzpicture}
\definecolor{ace15color}{RGB}{174,199,232}
\definecolor{heartcolor}{RGB}{255,127,14}
\definecolor{minimaxcolor}{RGB}{255,187,120}
\definecolor{mureka9color}{RGB}{152,223,138}
\definecolor{suno55color}{RGB}{214,39,40}
\definecolor{suno5color}{RGB}{255,152,150}
\begin{axis}[
    width=0.88\linewidth,
    height=0.42\linewidth,
    xlabel={Number of target samples $N$},
    ylabel={Target F1-score (\%)},
    xmin=0, xmax=55,
    ymin=79, ymax=100.5,
    xtick={0,5,10,20,50},
    ytick={80,85,90,95,100},
    axis line style={draw=black!70},
    tick style={draw=black!70},
    grid=none,
    legend style={font=\scriptsize, draw=none, fill=none, at={(0.5,-0.22)}, anchor=north, legend columns=3},
    tick label style={font=\scriptsize},
    label style={font=\small},
    every axis plot/.append style={line width=1.05pt},
]
\addplot+[ace15color, mark=o] coordinates {(0,90.23) (5,97.33) (10,99.36) (20,99.73) (50,99.73)};
\addlegendentry{ACE-Step 1.5}
\addplot+[heartcolor, mark=square*] coordinates {(0,97.10) (5,99.47) (10,99.42) (20,99.47) (50,99.73)};
\addlegendentry{HeartMuLa}
\addplot+[minimaxcolor, mark=triangle*] coordinates {(0,90.23) (5,98.22) (10,99.03) (20,99.61) (50,98.26)};
\addlegendentry{MiniMax 2.6}
\addplot+[mureka9color, mark=diamond*] coordinates {(0,95.13) (5,98.26) (10,99.02) (20,99.09) (50,99.35)};
\addlegendentry{Mureka V9}
\addplot+[suno55color, mark=star] coordinates {(0,90.88) (5,98.94) (10,99.47) (20,99.12) (50,98.77)};
\addlegendentry{Suno v5.5}
\addplot+[suno5color, mark=pentagon*] coordinates {(0,80.21) (5,96.68) (10,97.57) (20,97.59) (50,98.64)};
\addlegendentry{Suno v5}
\end{axis}
\end{tikzpicture}
\caption{Few-shot target F1-score of Sofia-VAG-FT under different numbers of target-generator samples. Colors follow the generator palette used in the representation visualization. All models are initialized from Sofia-VAG (MoE), with audio encoders frozen during adaptation.}
\label{fig:fewshot_ncurve}
\vspace{-0.8em}
\end{figure}

Figure~\ref{fig:fewshot_ncurve} further analyzes the effect of the number of target samples. Sofia-VAG-FT requires only a few examples to obtain strong target-domain performance: with merely $N=5$ target samples, all six generators already reach at least \textbf{96\%} F1-score. Increasing the number of samples to $N=10$ further brings most generators close to saturation around \textbf{99\%}. This trend shows that the adaptation is highly sample-efficient, making Sofia suitable for rapidly responding to newly emerging generators with minimal cost.

\section{Conclusion}
We present \textbf{Sofia}, a general and flexible framework for synthetic song detection that models music-intrinsic cues with feature-specific experts and adaptively fuses them via a Mixture-of-Experts module. We further introduce \textbf{MUSIC8K}, a new benchmark for evaluating detector generalization to synthetic songs from the latest music generators and robustness under realistic audio perturbations. Experiments demonstrate that Sofia learns generator-agnostic representations from complementary music features, achieving state-of-the-art generalization and strong robustness. Future work will extend Sofia with broader music features and stronger music understanding encoders.

\section*{Limitations}
Although \textbf{Sofia} provides a modular and scalable framework for synthetic song detection, this work mainly explores one representative instantiation, \textbf{Sofia-VAG}, based on vocal, audio-effect, and global musical-structure features. 
These features are effective for current generators, but they may not remain equally discriminative as AI music generation continues to improve. 
With the rapid iteration of music generators, any detector relying on a fixed set of encoders or feature assumptions may gradually become outdated. The value of Sofia is therefore not limited to the current Sofia-VAG model, but lies in its extensible architecture. As stronger music understanding encoders become available, they can be incorporated into Sofia to capture more subtle and complex musical cues.

\bibliographystyle{acl_natbib}
\bibliography{music}

\newpage
\appendix

\definecolor{tocblue}{RGB}{0,20,110}

\newcommand{\apptocsection}[3]{%
  \noindent
  {\color{tocblue}\bfseries\Large #1\hspace{1.0em}#2}%
  \hfill
  {\bfseries\Large #3}%
  \par\vspace{0.85em}
}

\newcommand{\apptocsubsection}[3]{%
  \noindent
  \hspace*{2.6em}%
  {\color{tocblue}\large #1\hspace{0.9em}#2}%
  \nobreak\leaders\hbox{\normalfont\large\hskip0.45em.\hskip0.45em}\hfill\nobreak
  {\large #3}%
  \par\vspace{0.45em}
}

\clearpage
\phantomsection
\addcontentsline{toc}{section}{Appendix Contents}

{\normalfont\Huge\scshape Table of Contents\par}
\vspace{2.8em}

\apptocsection{A}{Dataset Statistics}{\pageref{sec:Dataset Statistics}}
\apptocsubsection{A.1}{Training and Validation Splits}{\pageref{sec:training_validation_splits}}
\apptocsubsection{A.2}{Evaluation Sets}{\pageref{sec:evaluation_sets}}
\apptocsubsection{A.3}{Robustness Evaluation}{\pageref{sec:appendix_robustness}}

\vspace{0.8em}

\apptocsection{B}{Synthetic Song Generation Pipeline}{\pageref{sec:synthetic_generation_pipeline}}
\apptocsubsection{B.1}{Lyrics-to-Prompt Generation}{\pageref{sec:lyrics_to_prompt_generation}}
\apptocsubsection{B.2}{Music Synthesis}{\pageref{sec:music_synthesis}}

\vspace{0.8em}

\apptocsection{C}{Sofia Variants}{\pageref{sec:sofia_variants}}

\vspace{0.8em}

\apptocsection{D}{Baseline Details}{\pageref{sec:baseline_details}}
\apptocsubsection{D.1}{Spectrogram-Based Baselines}{\pageref{sec:spectrogram_based_baselines}}
\apptocsubsection{D.2}{Speech and Representation-Based Baselines}{\pageref{sec:speech_representation_based_baselines}}

\vspace{0.8em}

\apptocsection{E}{Music Feature Modeling}{\pageref{sec:music_feature_modeling}}
\apptocsubsection{E.1}{Vocal Features}{\pageref{sec:vocal_features}}
\apptocsubsection{E.2}{Audio-Effect Features}{\pageref{sec:audio_effect_features}}
\apptocsubsection{E.3}{Global Musical-Structure Features}{\pageref{sec:global_musical_structure_features}}

\vspace{0.8em}

\apptocsection{F}{Implementation Details}{\pageref{sec:implementation_details}}
\apptocsubsection{F.1}{Training Configuration}{\pageref{sec:training_configuration}}
\apptocsubsection{F.2}{Audio Preprocessing}{\pageref{sec:audio_preprocessing}}
\apptocsubsection{F.3}{Encoder Configuration}{\pageref{sec:encoder_configuration}}
\apptocsubsection{F.4}{Feature Extraction}{\pageref{sec:feature_extraction}}
\apptocsubsection{F.5}{Network Architecture}{\pageref{sec:network_architecture}}
\apptocsubsection{F.6}{Training Strategy}{\pageref{sec:training_strategy}}

\vspace{0.8em}

\apptocsection{G}{Additional Expert Weight Distributions}{\pageref{sec:additional_expert_weights}}

\newpage

\section{Dataset Statistics}
\label{sec:Dataset Statistics}

\subsection{Training and validation splits.}
\label{sec:training_validation_splits}
Table~\ref{tab:split_counts} summarizes the number of files per source in the SONICS training and validation splits used in our experiments.

\begin{table}[htbp]
\centering
\small
\begin{tabular}{llr}
\toprule
\textbf{Split} & \textbf{Source} & \textbf{Count} \\
\midrule
Train & Human & 10{,}797 \\
Train & Suno v3.5 & 9{,}017 \\
Train & Udio & 8{,}986 \\
\midrule
Valid & Human & 1{,}203 \\
Valid & Suno v3.5 & 983 \\
Valid & Udio & 1{,}014 \\
\bottomrule
\end{tabular}
\caption{SONICS training and validation splits.}
\label{tab:split_counts}
\end{table}

\subsection{Evaluation sets.}
\label{sec:evaluation_sets}
Table~\ref{tab:dataset_source_counts} summarizes the per-source file counts of MUSIC8K-O, MoM, and SONICS used for evaluation.

\begin{table}[htbp]
\centering
\small
\setlength{\tabcolsep}{6pt}
\renewcommand{\arraystretch}{1.08}
\begin{tabular}{llr}
\toprule
\textbf{Dataset} & \textbf{Source / Generator} & \textbf{Count} \\
\midrule
\multirow{8}{*}{MUSIC8K-O}
& ACE-Step 1.0 & 1{,}656 \\
& ACE-Step 1.5 & 989 \\
& HeartMuLa & 989 \\
& Minimax 2.6 & 304 \\
& Mureka O1 & 213 \\
& Mureka V9 & 662 \\
& Suno v5 & 400 \\
& Suno v5.5 & 330 \\
\midrule
\multirow{6}{*}{MoM}
& Riffusion & 7{,}057 \\
& DiffRhythm & 4{,}606 \\
& Yue & 5{,}278 \\
& Suno v3 & 3{,}512 \\
& Suno v3.5 & 23{,}695 \\
& Suno v4 & 48 \\
\midrule
\multirow{5}{*}{SONICS}
& Suno v2 & 2{,}084 \\
& Suno v3 & 4{,}285 \\
& Suno v3.5 & 19{,}057 \\
& Udio 32 & 4{,}903 \\
& Udio 130 & 18{,}745 \\
\bottomrule
\end{tabular}
\caption{Per-source file counts of MUSIC8K-O, MoM, and SONICS used in our evaluation.}
\label{tab:dataset_source_counts}
\end{table}

\subsection{Robustness Evaluation}
\label{sec:appendix_robustness}

\paragraph{Audio perturbations.}
MUSIC8K-P is constructed from a subset of MUSIC8K-O by applying one of three audio transformations:
\begin{itemize}
  \item \textbf{Pitch shift.} Random semitone shift $n \sim \mathcal{U}[-2, 2]$ using a phase-vocoder-based pitch shift.
  \item \textbf{Time stretch.} Random time-stretch factor $r \sim \mathcal{U}[0.8, 1.2]$, where $r<1$ slows down the audio and $r>1$ speeds it up.
  \item \textbf{Additive noise.} White Gaussian noise is added to achieve an SNR of 20 dB, and the result is clipped to $[-1,1]$ to avoid overflow.
\end{itemize}

\paragraph{Robustness dataset composition.}
All three perturbed splits share the same source composition. Each split contains 2{,}625 songs, resulting in 7{,}875 perturbed songs in total. Table~\ref{tab:robustness} reports the per-source counts for each perturbation split.

\begin{table}[htbp]
\centering
\small
\begin{tabular}{lr}
\toprule
\textbf{Source} & \textbf{Count} \\
\midrule
ACE-Step 1.5 & 989 \\
Minimax 2.6 & 304 \\
Mureka V9 & 662 \\
Suno v5.5 & 330 \\
Suno v5 & 340 \\
\bottomrule
\end{tabular}
\caption{Per-source file counts in each MUSIC8K-P perturbation split, including pitch shift, time stretch, and additive noise.}
\label{tab:robustness}
\end{table}

\section{Synthetic Song Generation Pipeline}
\label{sec:synthetic_generation_pipeline}

\subsection{Lyrics-to-prompt generation.}
\label{sec:lyrics_to_prompt_generation}
The pipeline starts from a large lyrics corpus\footnote{\url{https://www.kaggle.com/datasets/carlosgdcj/genius-song-lyrics-with-language-information}}.
A preprocessing script filters and cleans the data, selects a pop-only subset, and keeps one song per artist to form a 1,000-song seed set.
Qwen-Audio~\cite{Chu2023QwenAudioAU} is then used to extract musical attributes and produce a compact text-to-music prompt in JSON format, including genre or mood description, BPM, duration, key scale, language, and time signature.

\subsection{Music synthesis.}
\label{sec:music_synthesis}
The generated prompts are used to call music generation APIs, which produce the final synthetic songs. Based on the Artificial Analysis music generation leaderboard\footnote{\url{https://artificialanalysis.ai/music/leaderboard/vocals}}, MUSIC8K-O is constructed from recent commercial and open-source generators, including ACE-Step 1.0, ACE-Step 1.5, Mureka O1, Mureka V9, Suno v5, Suno v5.5, Minimax 2.6, and HeartMuLa.

\section{Sofia Variants.}
\label{sec:sofia_variants}
We also evaluate different Sofia feature instantiations. 
Sofia-V1(Wav2Vec2) and Sofia-V1(RawNet2) use vocal features, Sofia-A1(FxPP) uses audio-effect features, and Sofia-G1(MuQ) and Sofia-G1(MERT) use global musical-structure features. 
Sofia-G2(MuQ+MERT) combines two global music encoders using MoE fusion. 
Sofia-VAG (concat) and Sofia-VAG (MoE) use the same vocal, audio-effect, and global feature set, but differ in the fusion strategy. 
Note that RawNet2 and Sofia-V1(RawNet2) are different settings: RawNet2 is a standalone vocal-track binary classifier, whereas Sofia-V1(RawNet2) uses the pretrained RawNet2 model as a frozen encoder within the Sofia framework and trains the Sofia projection, fusion, and classification modules.

\section{Baseline Details}
\label{sec:baseline_details}

We compare Sofia with several representative baselines from prior SSD studies.

\subsection{Spectrogram-based baselines.}
\label{sec:spectrogram_based_baselines}
We include ConvNeXt~\cite{Liu2022ACF}, ViT~\cite{Dosovitskiy2020AnII}, EfficientViT~\cite{Liu2023EfficientViTME}, and three SpecTTTra variants, i.e., SpecTTTra-$\gamma$, SpecTTTra-$\beta$, and SpecTTTra-$\alpha$~\cite{rahman2024sonics}. 
These methods convert audio into spectrogram representations and formulate synthetic song detection as an image classification or spectro-temporal modeling problem. 
ConvNeXt is a CNN-based image classification backbone with strong local inductive bias. 
ViT treats spectrogram patches as visual tokens and models their global dependencies through self-attention. 
EfficientViT is a memory-efficient vision transformer designed to reduce the computational cost of standard vision transformers. 
SpecTTTra is a spectro-temporal transformer designed for synthetic song detection, where its $\gamma$, $\beta$, and $\alpha$ variants correspond to different model scales.

\subsection{Speech and representation-based baselines.}
\label{sec:speech_representation_based_baselines}
RawNet2~\cite{tak2021end} is included as a speech anti-spoofing baseline. 
In our experiments, RawNet2 denotes a binary classifier trained on the vocal tracks from the SONICS training split. 
CLAM~\cite{batramelody} is selected as a strong state-of-the-art SSD baseline. 
It uses MERT and Wav2Vec2 representations in a dual-stream contrastive framework to model the consistency between music and vocal-related representations.

\section{Music Feature Modeling}
\label{sec:music_feature_modeling}

To model multi-dimensions musical features, Sofia-VAG uses five feature-specific encoders covering vocal, audio-effect, and global musical-structure features.

\subsection{Vocal features.}
\label{sec:vocal_features}
Wav2Vec2~\cite{baevski2020wav2vec} is a self-supervised speech representation model that captures temporal dependencies and phonetic structures in vocal signals. In Sofia-VAG, it is used to model vocal-related cues in singing voices. RawNet2~\cite{tak2021end} is originally designed for speech anti-spoofing, making it useful for modeling vocal naturalness and authenticity.

\subsection{Audio-effect features.}
\label{sec:audio_effect_features}
Fx-Encoder++~\cite{yeh2025fx} is designed to extract effect-aware representations from audio mixtures. It provides embeddings that reflect acoustic environment characteristics, spatial cues, and production-related artifacts.

\subsection{Global musical-structure features.}
\label{sec:global_musical_structure_features}
For global features, we use MuQ~\cite{zhu2025muq} and MERT~\cite{yizhi2023mert}. These self-supervised music representation models produce high-level embeddings that capture semantic and structural properties of music, such as song organization, harmony, rhythm, and overall composition patterns.

\section{Implementation Details}
\label{sec:implementation_details}

\subsection{Training configuration.}
\label{sec:training_configuration}
All experiments are conducted on a single NVIDIA H800 GPU.
Sofia-VAG is trained with AdamW using a learning rate of $1\times10^{-3}$, weight decay of $1\times10^{-4}$, and betas $(0.9, 0.98)$.
The maximum gradient norm is set to $5.0$.
We train the model for $1$ epoch with a batch size of $32$.
Automatic mixed precision is disabled by default.

\subsection{Audio preprocessing.}
\label{sec:audio_preprocessing}
Input audio is first loaded at a base sampling rate of $44.1$ kHz and normalized to unit peak.
Each sample is then cropped or padded to a fixed segment length.
For each branch, the waveform is resampled to the target sampling rate required by the corresponding encoder.
Branches configured with \texttt{use\_vocals=true} use the separated vocal waveform, while the remaining branches use the full-song waveform.
In Sofia-VAG, the RawNet2 branch uses vocal inputs, while the other branches operate on full-song audio.

\subsection{Encoder configuration.}
\label{sec:encoder_configuration}
Sofia-VAG uses five encoders to instantiate three groups of music-intrinsic features.
The encoder metadata is summarized in Table~\ref{tab:encoder_metadata}.

\begin{table}[H]
\centering
\footnotesize
\setlength{\tabcolsep}{3pt}
\renewcommand{\arraystretch}{1.05}
\begin{tabular}{lllll}
\toprule
\textbf{Encoder} & \textbf{Feature group} & \textbf{Target SR} & \textbf{Ch.} & \textbf{Dim.} \\
\midrule
Fx-Encoder++ & Audio-effect & 44{,}100 & 2 & 128 \\
MuQ & Global & 24{,}000 & 1 & 1024 \\
MERT & Global & 24{,}000 & 1 & 768 \\
Wav2Vec2 & Vocal & 16{,}000 & 1 & 768 \\
RawNet2 & Vocal & 16{,}000 & 1 & 1024 \\
\bottomrule
\end{tabular}
\caption{Encoder metadata in Sofia-VAG.}
\label{tab:encoder_metadata}
\end{table}

\subsection{Feature extraction.}
\label{sec:feature_extraction}
Each encoder extracts a fixed-dimensional representation from its corresponding preprocessed waveform.
For Fx-Encoder++, we use the 128-dimensional effect-aware embedding returned by the local wrapper, with stereo audio clamped to $[-1,1]$ before extraction.
For MuQ, we use \texttt{last\_hidden\_state} and apply temporal mean pooling to obtain a 1024-dimensional global music representation.
For MERT and Wav2Vec2, we extract hidden states, apply the selected pooling strategy, and average over time to obtain 768-dimensional embeddings.
For RawNet2, we use the 1024-dimensional pre-classification embedding.

\subsection{Network architecture.}
\label{sec:network_architecture}
Each encoder output is first $\ell_2$-normalized and projected into a shared 256-dimensional space:
\[
\begin{aligned}
\mathrm{Linear}(d_i,256) 
&\rightarrow \mathrm{LayerNorm}(256) &\rightarrow \mathrm{GELU} 
\rightarrow \mathrm{Dropout}(0.1),
\end{aligned}
\]
where $d_i$ denotes the output dimension of the $i$-th encoder.
The projected branch features are fused by the MoE-based feature fusion module.
The final prediction head is a linear softmax classifier, i.e., $\mathrm{Linear}(256,2)$.

\subsection{Training strategy.}
\label{sec:training_strategy}
Sofia follows a two-stage training strategy.
In Stage I, feature-specific encoders can be trained or adapted with their own objectives.
In the Sofia-VAG implementation, only the RawNet2 encoder is trained in this stage to better capture vocal-related authenticity cues, while Fx-Encoder++, MuQ, MERT, and Wav2Vec2 are initialized from pretrained checkpoints and kept fixed.
In Stage II, all encoder backbones are frozen.
The framework then trains the branch projectors, MoE-based fusion module, output projection, and classification head using softmax cross-entropy.

\begin{algorithm}[H]
\footnotesize
\caption{Two-Stage Training of Sofia}
\label{alg:sofia_training}
\begin{algorithmic}[1]
\Require Training dataset $\mathcal{D}=\{(x_i, y_i)\}_{i=1}^{M}$
\Require Encoders $\{E_k\}_{k=1}^{n}$, projectors $\{P_k\}_{k=1}^{n}$
\Require Experts $\{\psi_k\}_{k=1}^{n}$, gating network $G_\theta$, output projection $O$, classifier $(W_c,b_c)$

\State \textbf{Stage I: Encoder Training}
\State Train or adapt selected encoders with feature-specific objectives; keep the remaining pretrained encoders fixed.

\State \textbf{Stage II: Framework Training}
\State Freeze all encoders $\{E_k\}_{k=1}^{n}$.

\For{each mini-batch $\{(x_i,y_i)\}_{i=1}^{B}$}
    \For{each sample $(x_i,y_i)$ in the mini-batch}
        \For{$k=1$ to $n$}
            \State $f_k = P_k\!\left(\frac{E_k(x_i)}{\|E_k(x_i)\|_2+\epsilon}\right)$
        \EndFor
        \State $\mathbf{z} = \mathrm{Concat}(f_1,\ldots,f_n)$
        \State $\alpha = \mathrm{Softmax}(G_\theta(\mathbf{z}))$
        \State $\mathbf{h} = \sum_{k=1}^{n}\alpha_k\psi_k(f_k) + \frac{1}{n}\sum_{k=1}^{n} f_k$
        \State $p = \mathrm{Softmax}\!\left(W_c \frac{O(\mathbf{h})}{\|O(\mathbf{h})\|_2+\epsilon}+b_c\right)$
        \State Compute $\mathrm{CE}(p,y_i)$
    \EndFor
    \State Update $\{P_k\}$, $\{\psi_k\}$, $G_\theta$, $O$, and $(W_c,b_c)$ with the averaged mini-batch loss.
\EndFor
\end{algorithmic}
\end{algorithm}


\section{Additional Expert Weight Distributions}
\label{sec:additional_expert_weights}

This section provides supplementary dataset-level expert weight distributions for different generators.

\begin{figure}[htbp]
  \centering
  \includegraphics[width=0.9\columnwidth]{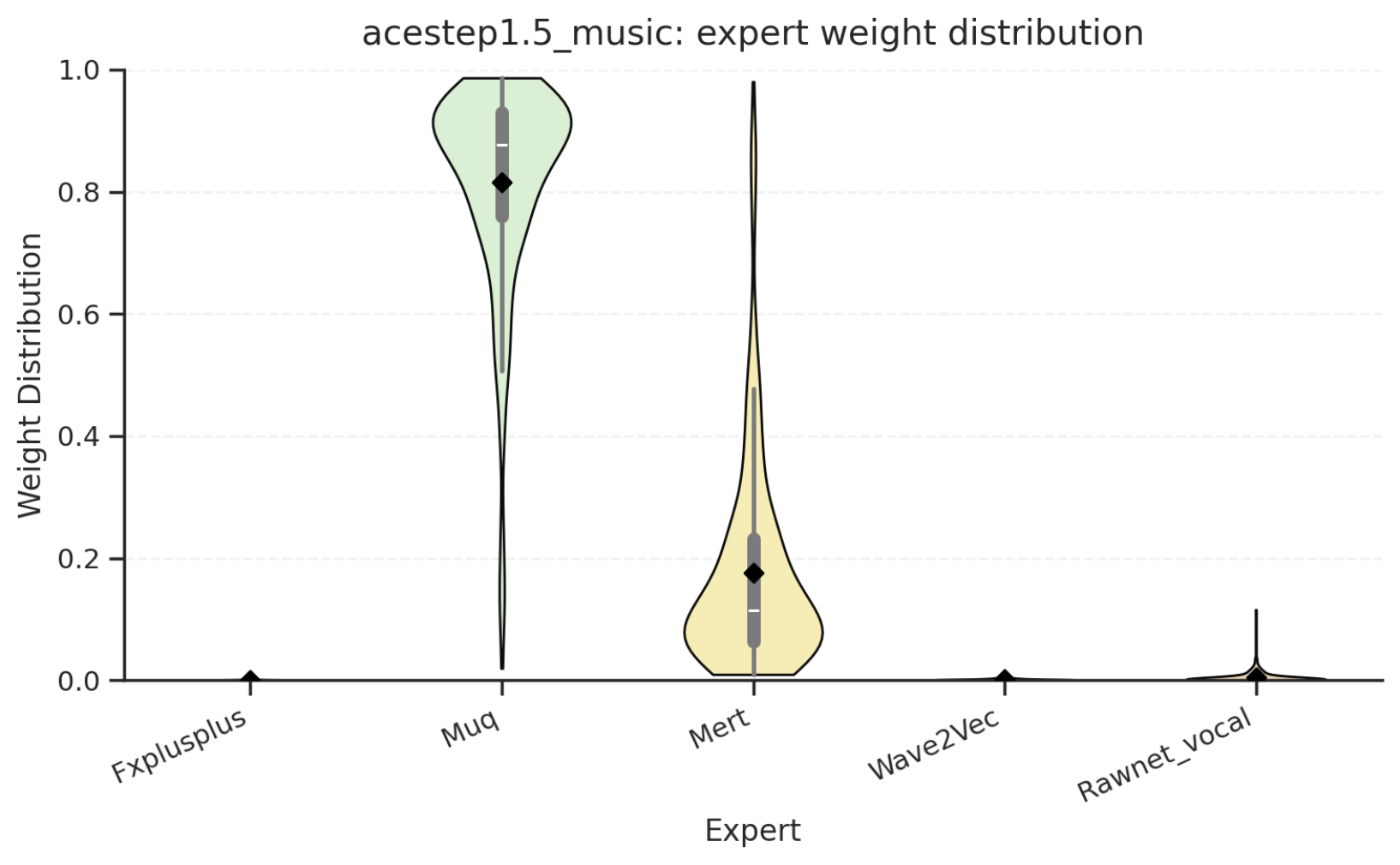}
  \caption{Expert weight distribution for ACE-Step 1.5.}
  \label{fig:weight_acestep15}
\end{figure}

\begin{figure}[htbp]
  \centering
  \includegraphics[width=0.9\columnwidth]{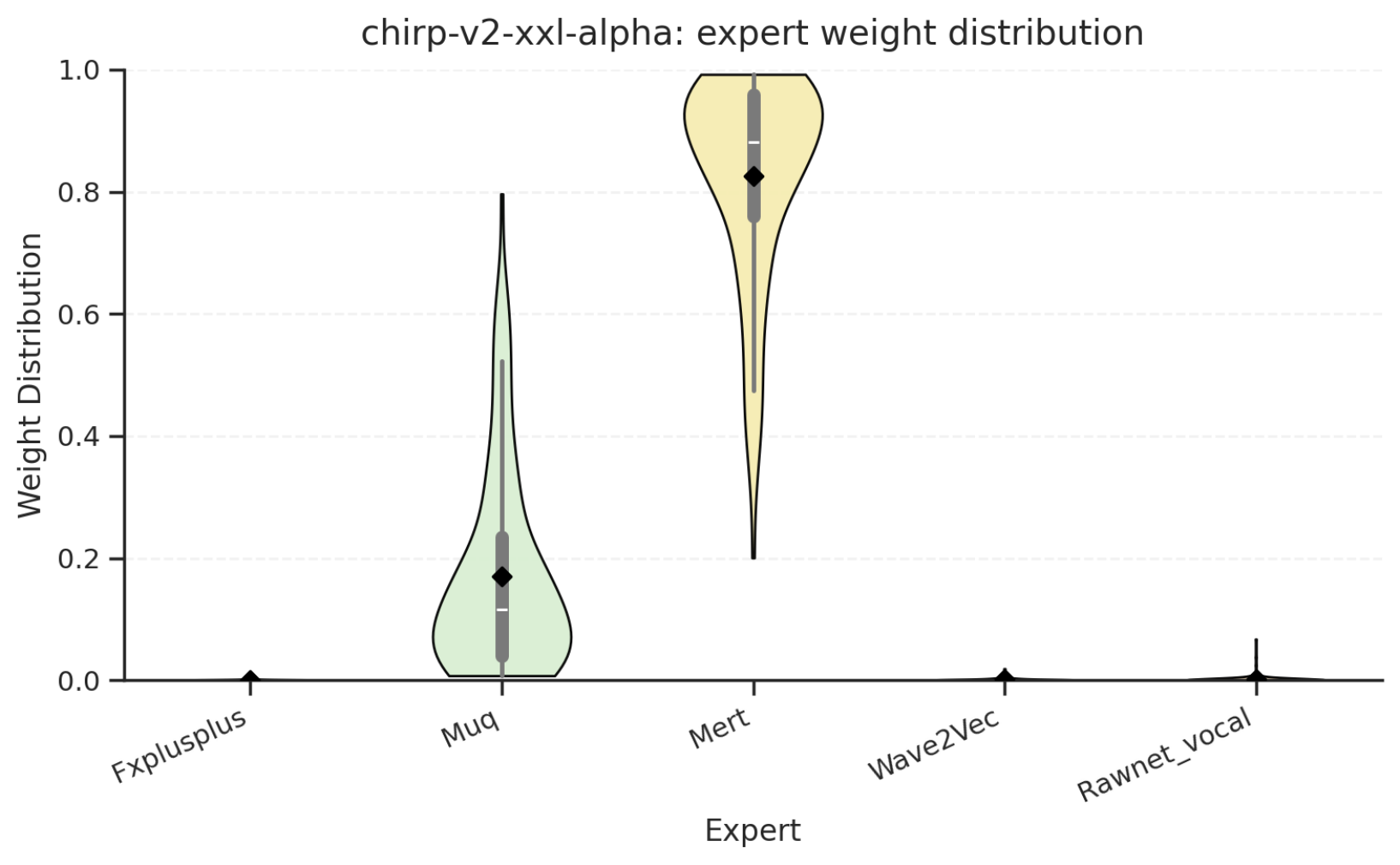}
  \caption{Expert weight distribution for Suno v2.}
  \label{fig:weight_suno_v2_xxl_alpha}
\end{figure}

\begin{figure}[htbp]
  \centering
  \includegraphics[width=0.9\columnwidth]{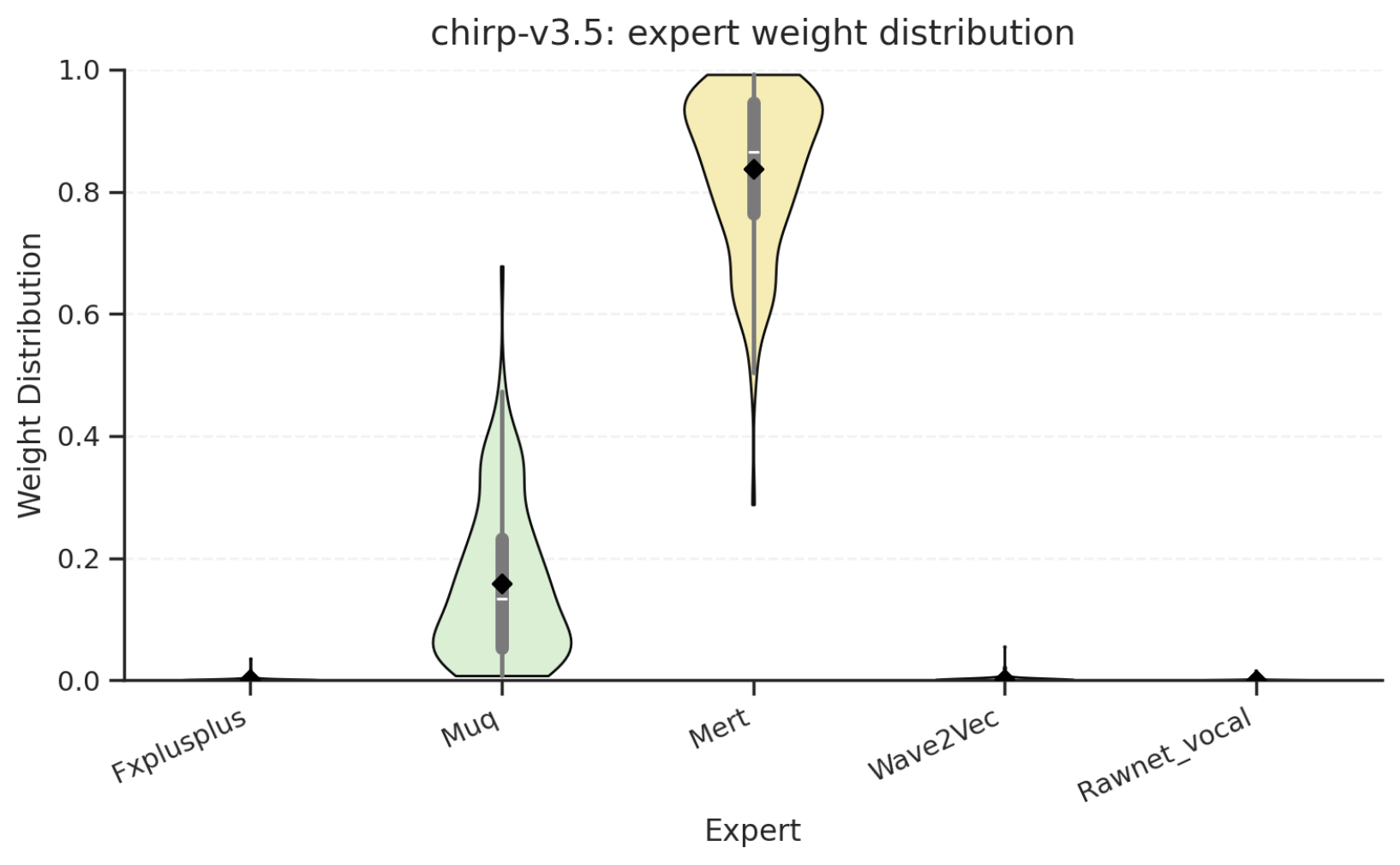}
  \caption{Expert weight distribution for Suno v3.5.}
  \label{fig:weight_suno_v35}
\end{figure}

\begin{figure}[htbp]
  \centering
  \includegraphics[width=0.9\columnwidth]{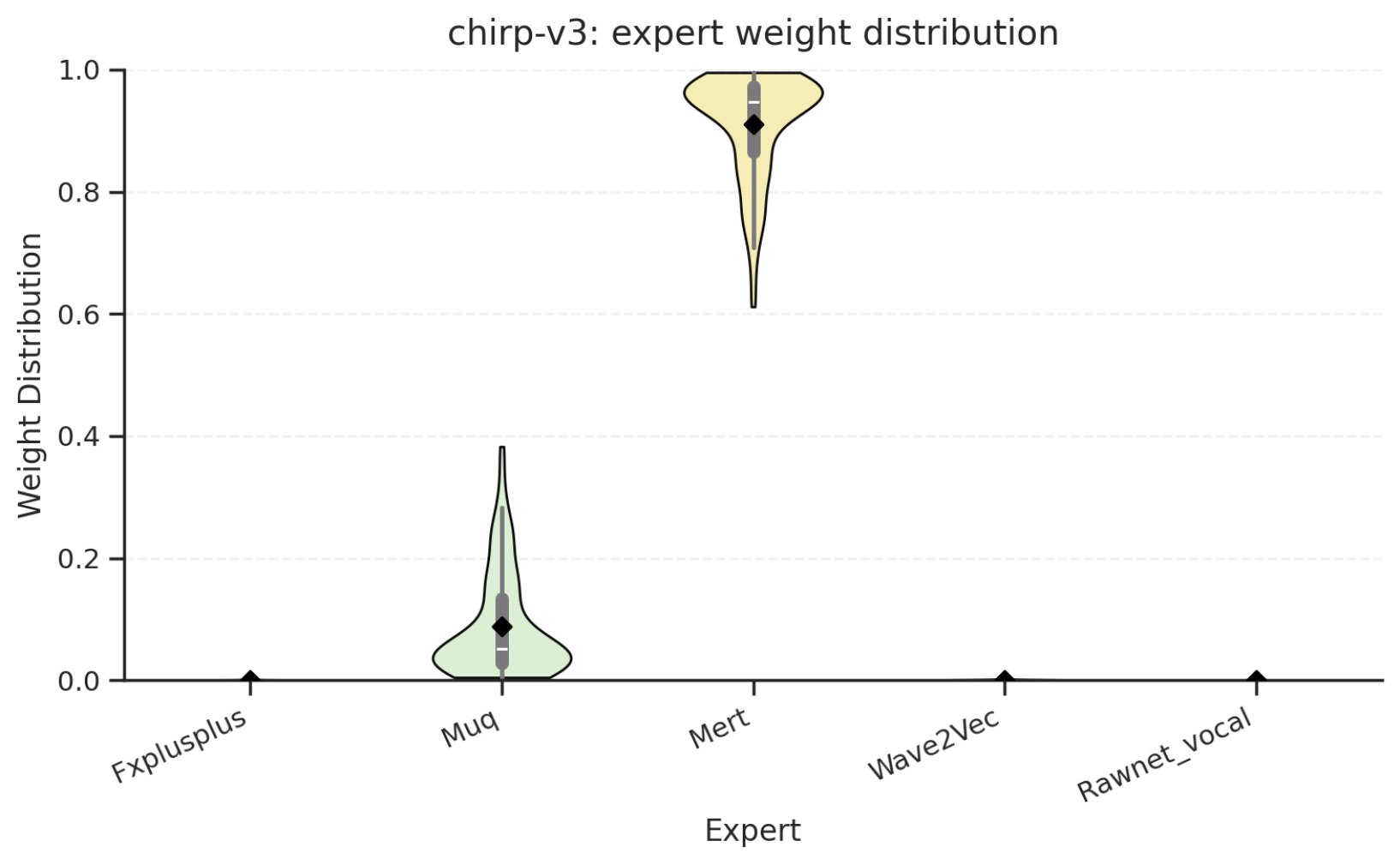}
  \caption{Expert weight distribution for Suno v3.}
  \label{fig:weight_suno_v3}
\end{figure}

\begin{figure}[htbp]
  \centering
  \includegraphics[width=0.9\columnwidth]{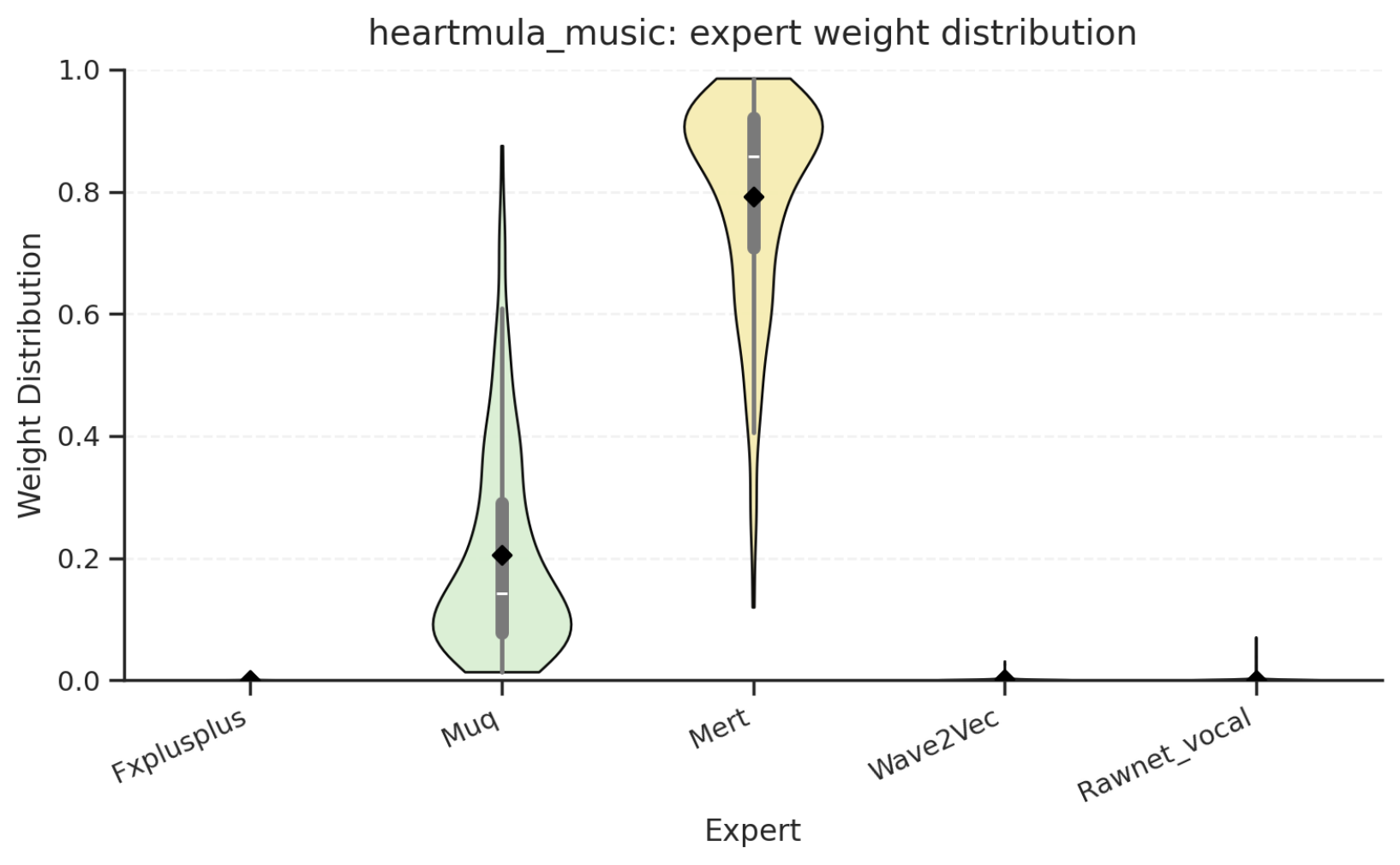}
  \caption{Expert weight distribution for HeartMuLa.}
  \label{fig:weight_heartmula}
\end{figure}

\begin{figure}[htbp]
  \centering
  \includegraphics[width=0.9\columnwidth]{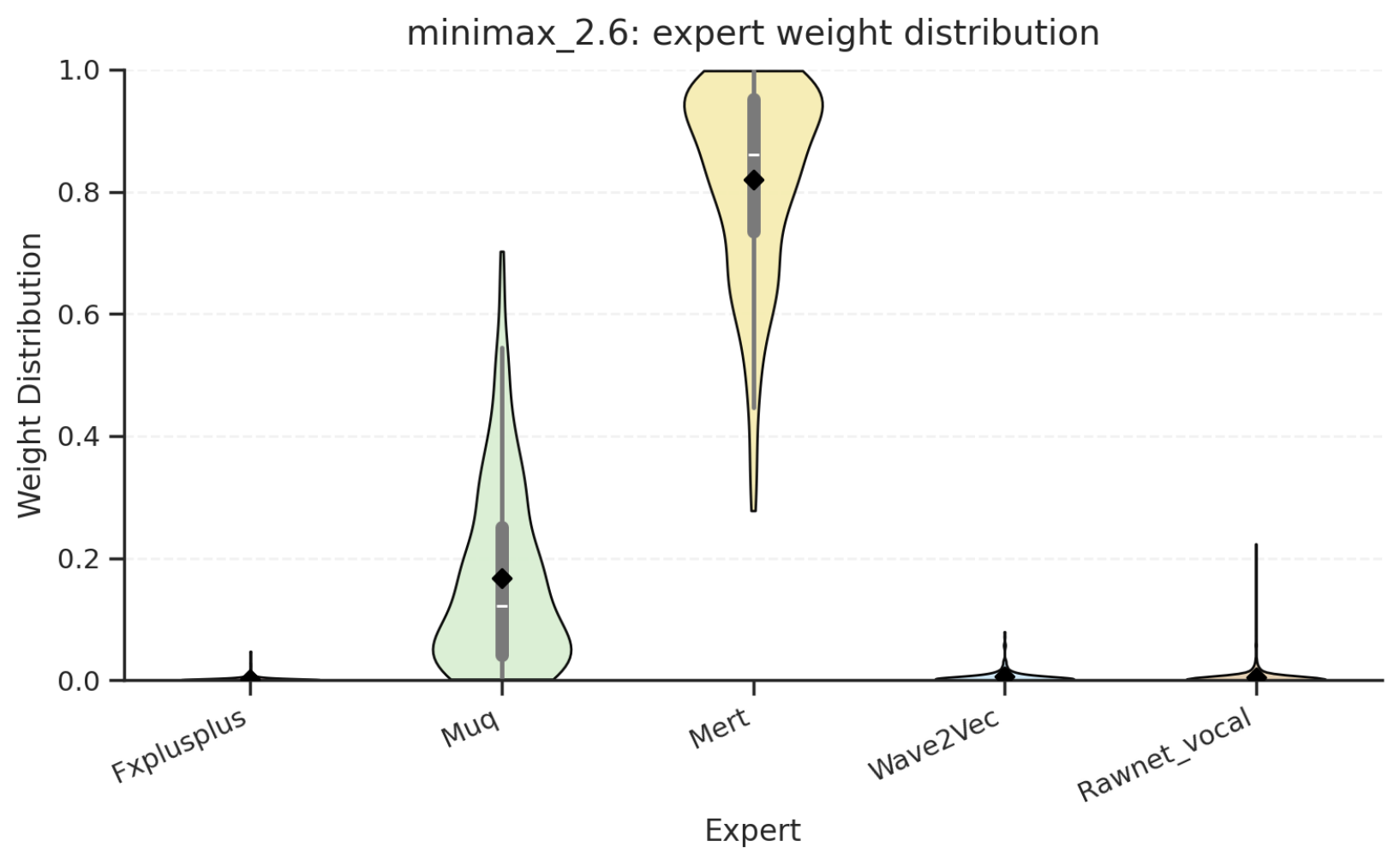}
  \caption{Expert weight distribution for Minimax 2.6.}
  \label{fig:weight_minimax26}
\end{figure}

\begin{figure}[htbp]
  \centering
  \includegraphics[width=0.9\columnwidth]{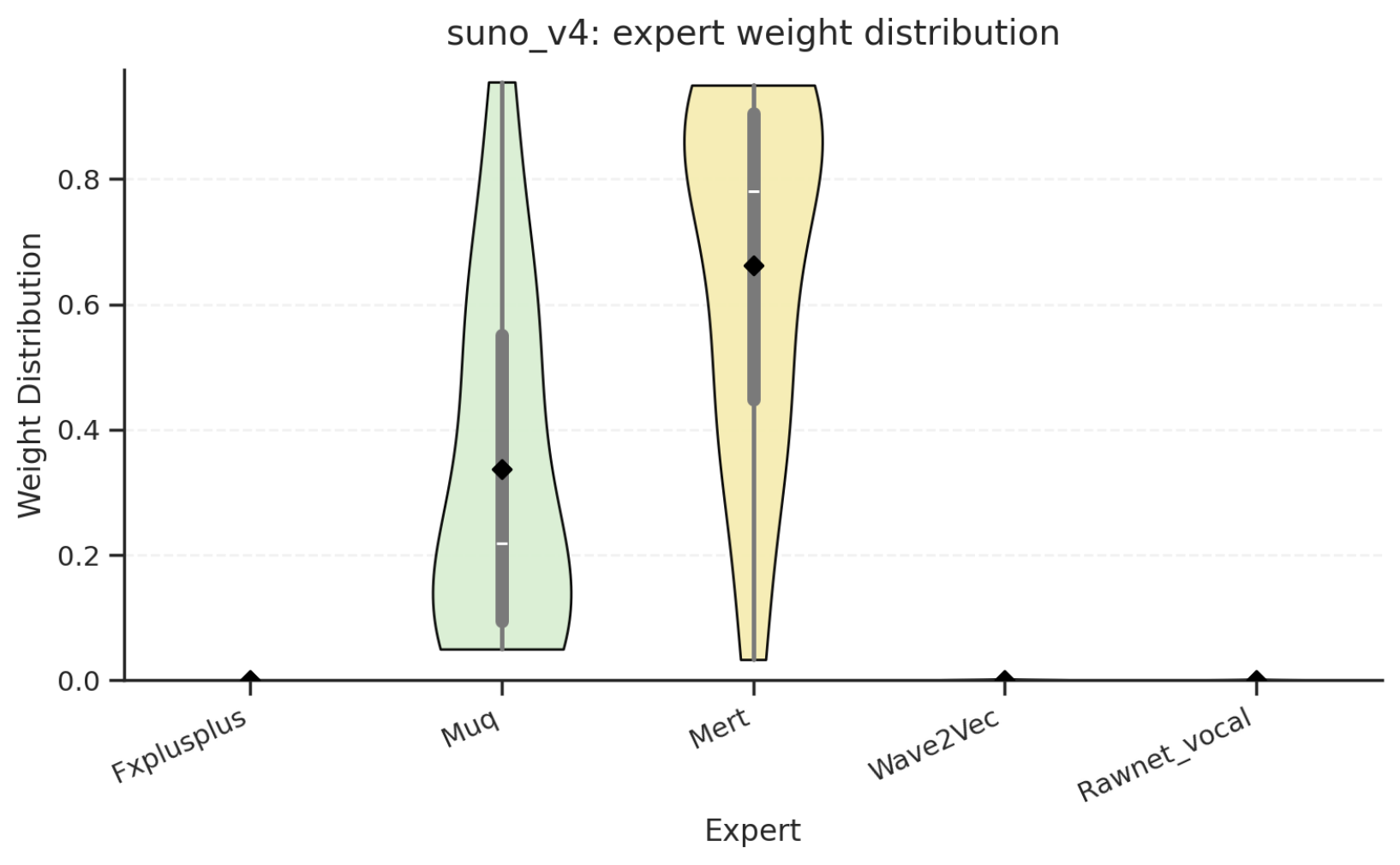}
  \caption{Expert weight distribution for Suno v4.}
  \label{fig:weight_suno_v4}
\end{figure}

\begin{figure}[htbp]
  \centering
  \includegraphics[width=0.9\columnwidth]{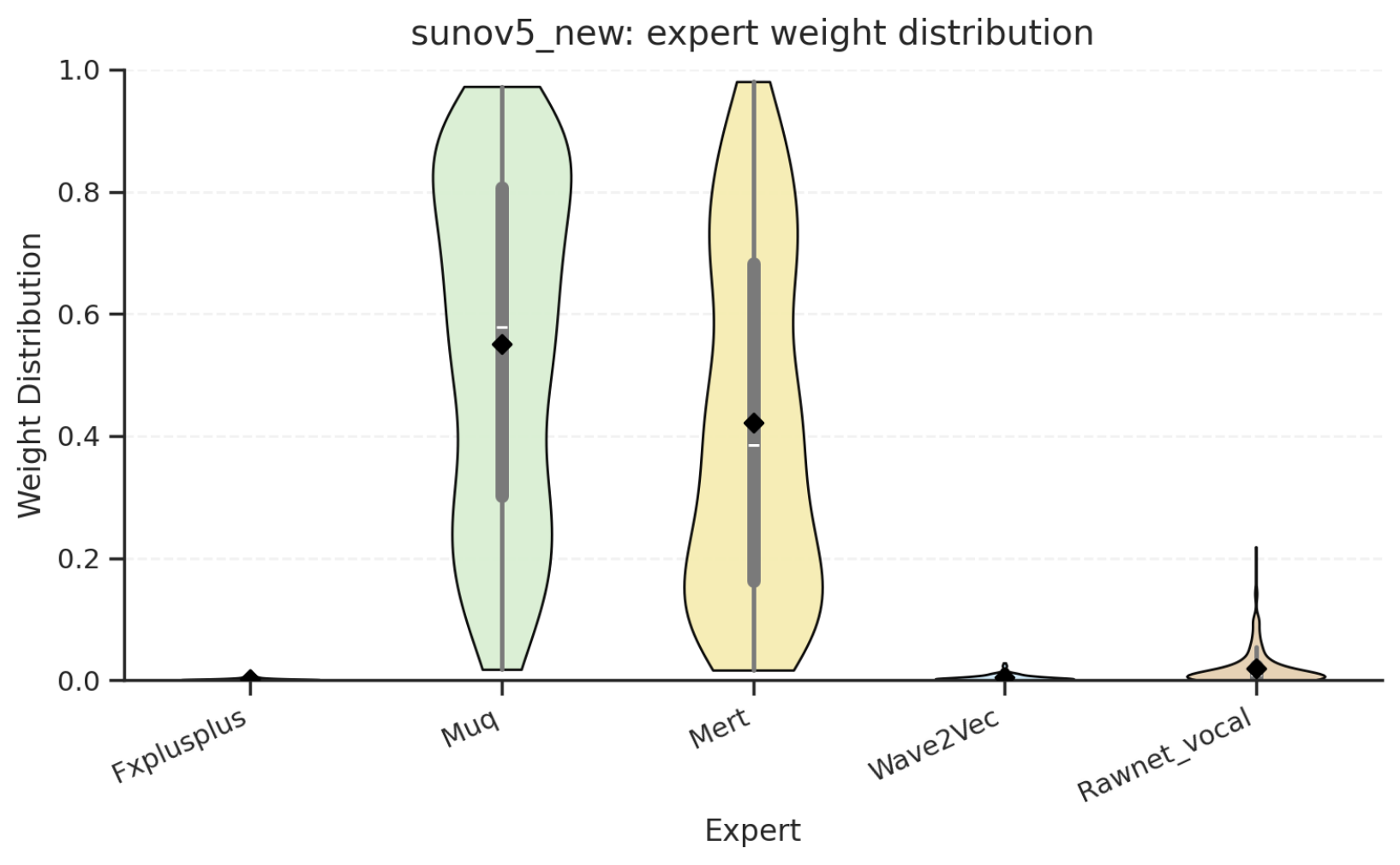}
  \caption{Expert weight distribution for Suno v5.}
  \label{fig:weight_suno_v5}
\end{figure}

\begin{figure}[htbp]
  \centering
  \includegraphics[width=0.9\columnwidth]{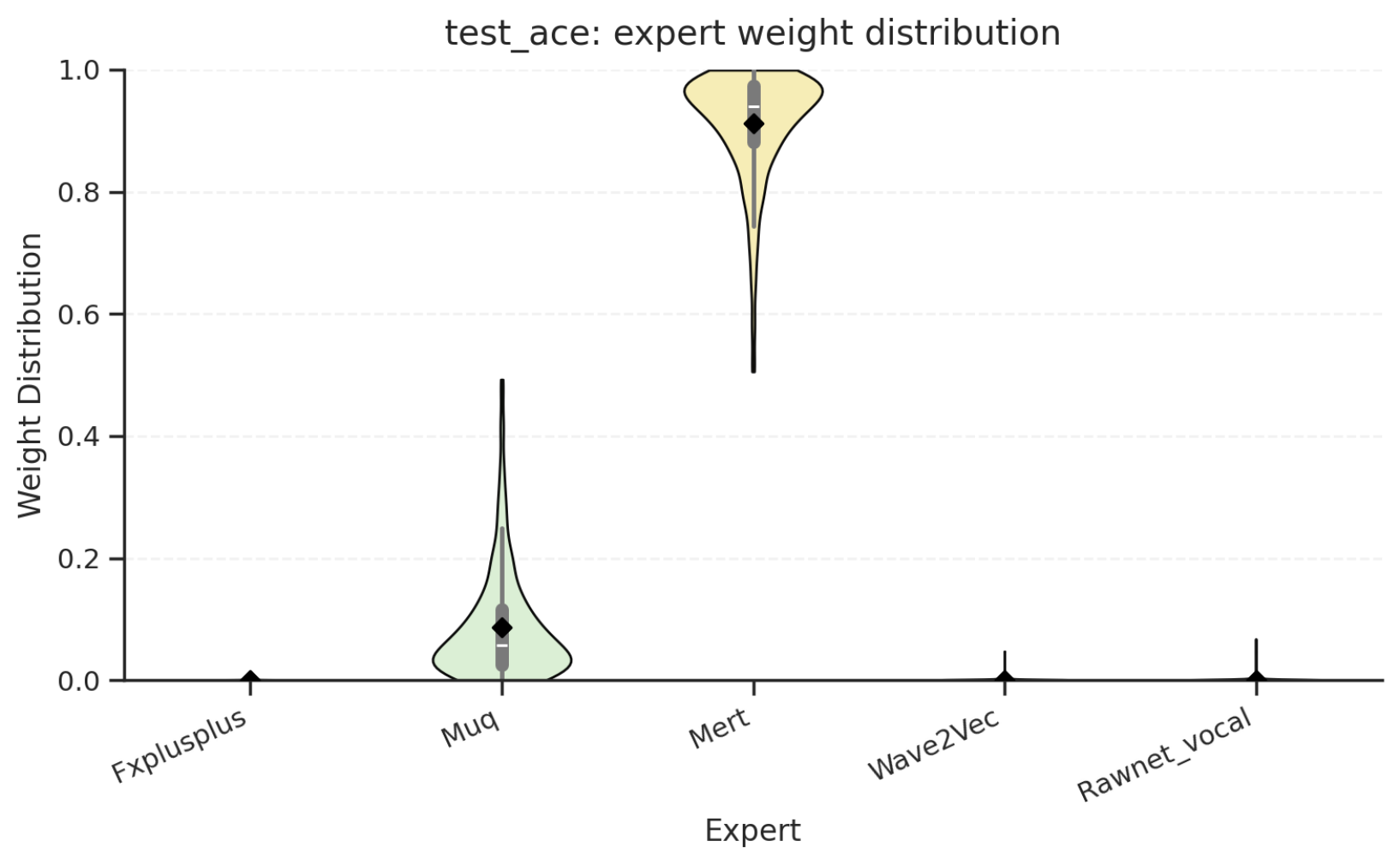}
  \caption{Expert weight distribution for ACE-Step 1.0.}
  \label{fig:weight_ace10}
\end{figure}

\begin{figure}[htbp]
  \centering
  \includegraphics[width=0.9\columnwidth]{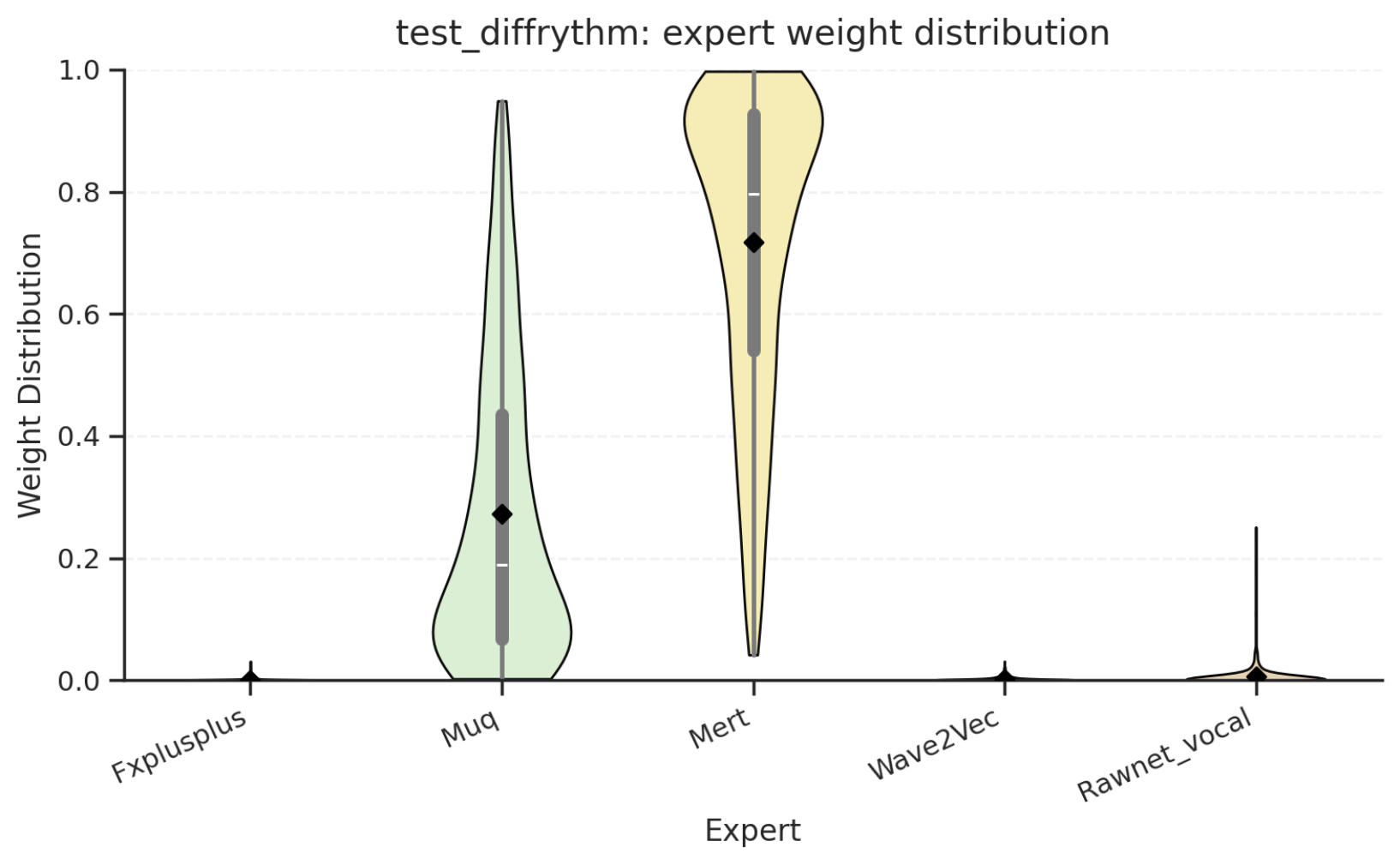}
  \caption{Expert weight distribution for DiffRhythm.}
  \label{fig:weight_diffrythm}
\end{figure}

\begin{figure}[htbp]
  \centering
  \includegraphics[width=0.9\columnwidth]{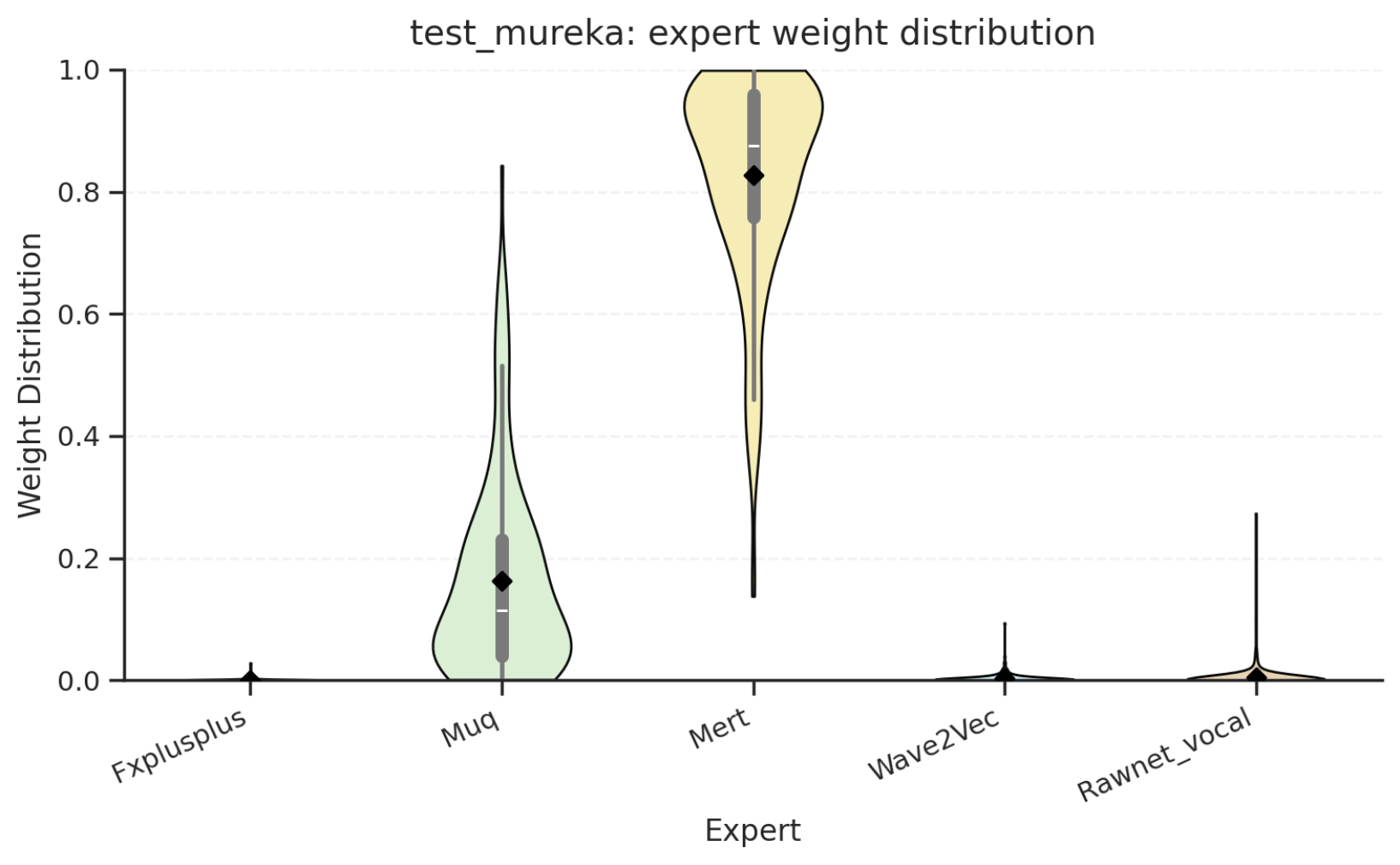}
  \caption{Expert weight distribution for Mureka v9.}
  \label{fig:weight_mureka_v9}
\end{figure}

\begin{figure}[htbp]
  \centering
  \includegraphics[width=0.9\columnwidth]{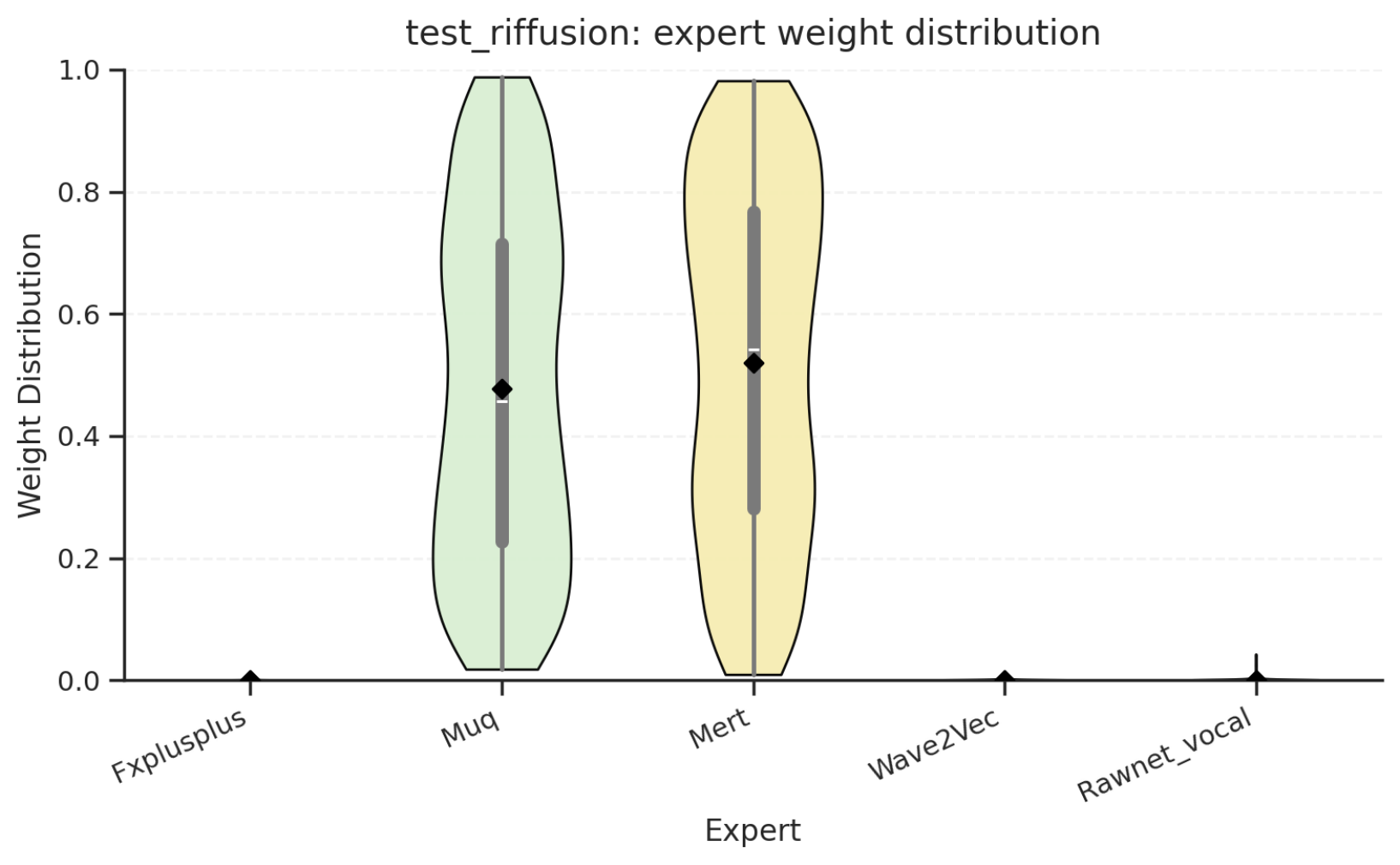}
  \caption{Expert weight distribution for Riffusion.}
  \label{fig:weight_riffusion}
\end{figure}

\begin{figure}[htbp]
  \centering
  \includegraphics[width=0.9\columnwidth]{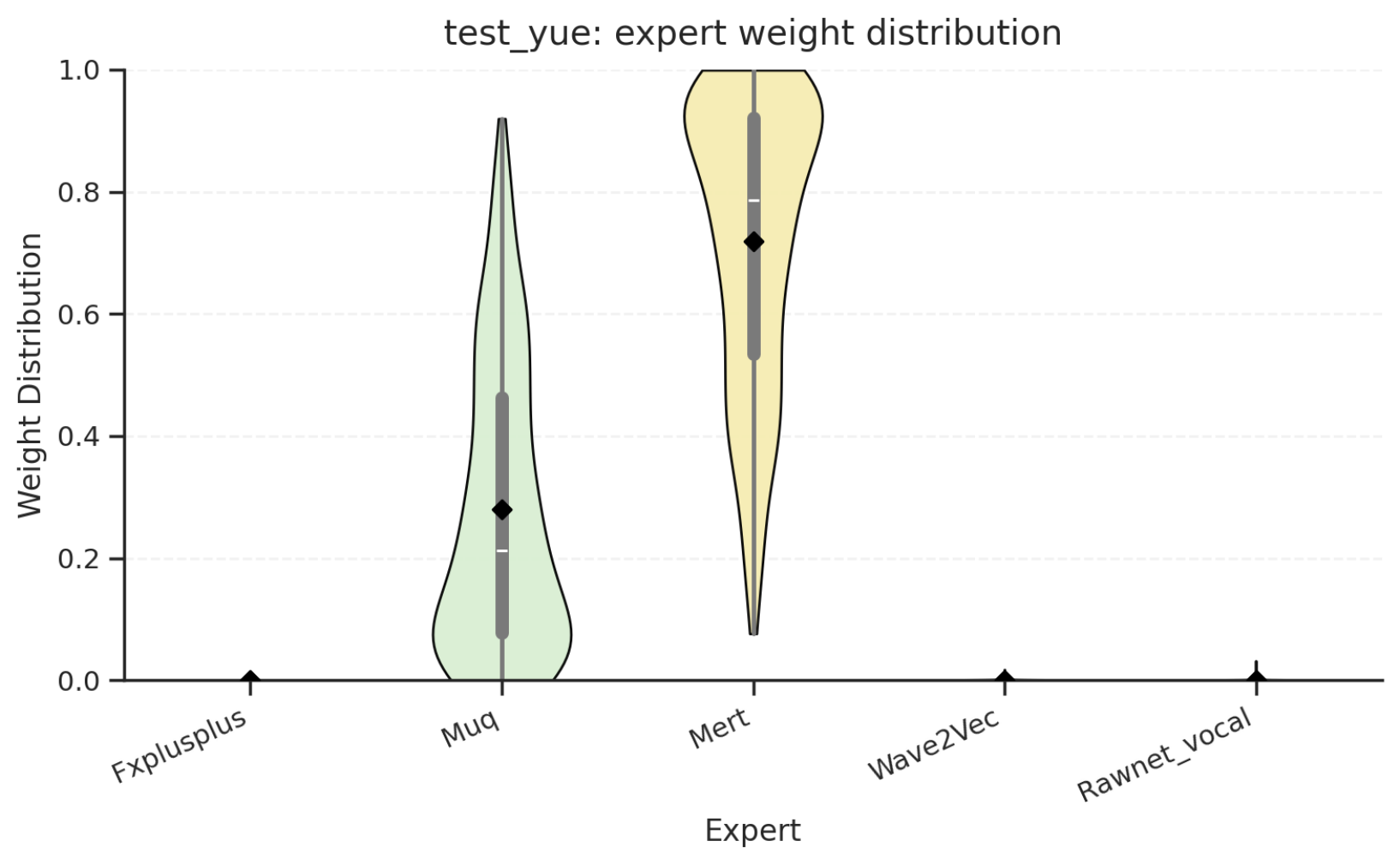}
  \caption{Expert weight distribution for Yue.}
  \label{fig:weight_yue}
\end{figure}

\begin{figure}[htbp]
  \centering
  \includegraphics[width=0.9\columnwidth]{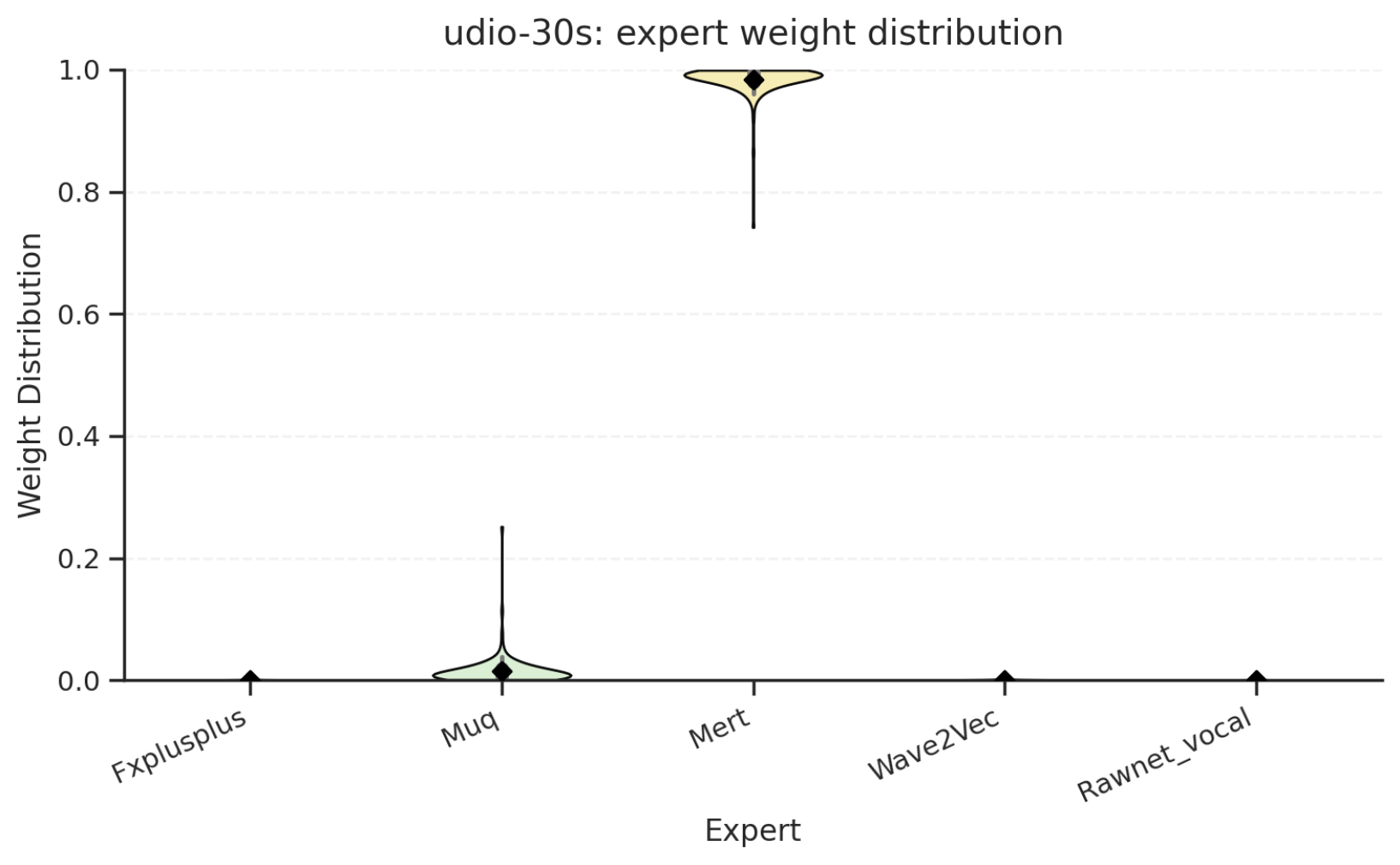}
  \caption{Expert weight distribution for Udio 32.}
  \label{fig:weight_udio32}
\end{figure}

\begin{figure}[htbp]
  \centering
  \includegraphics[width=0.9\columnwidth]{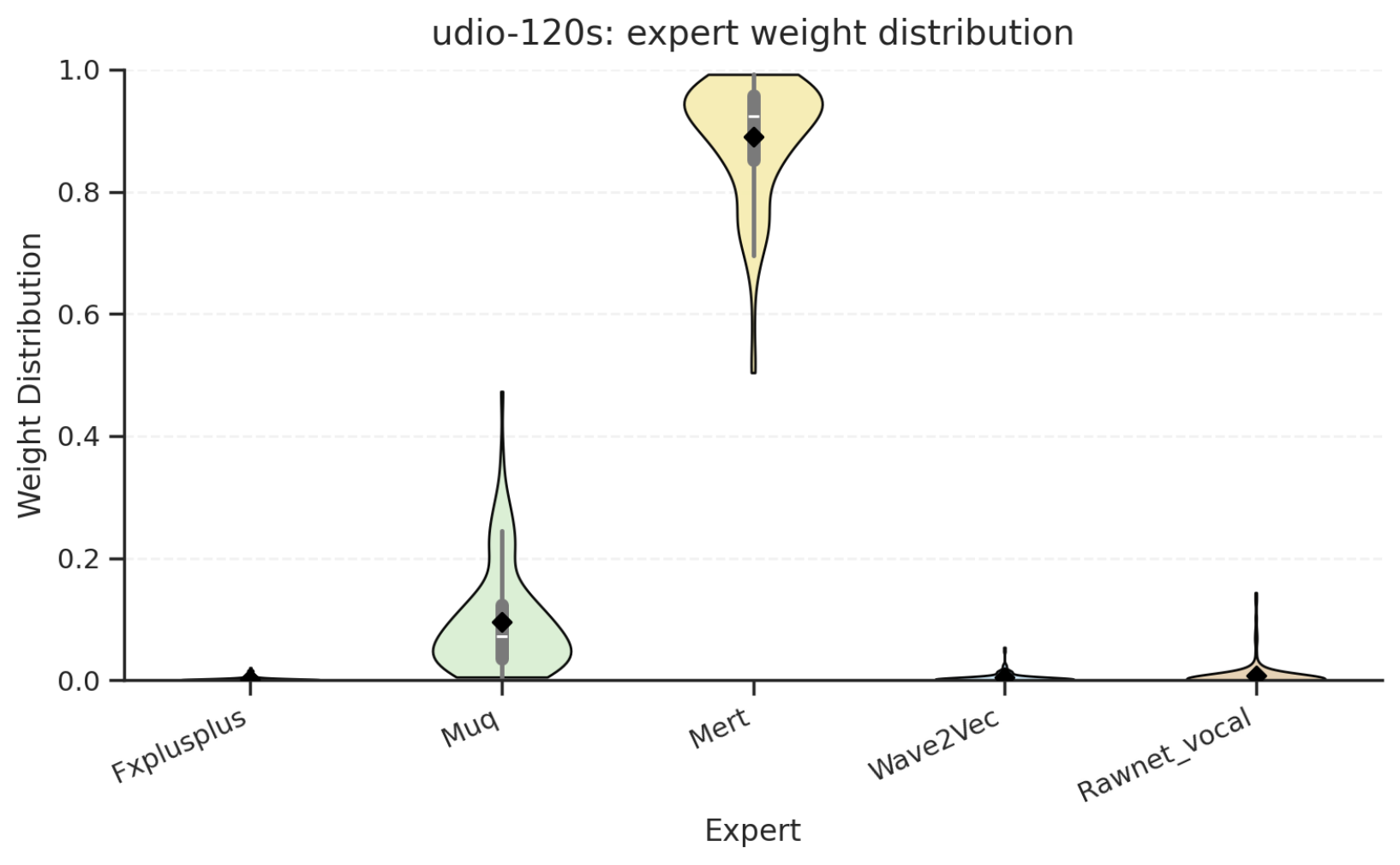}
  \caption{Expert weight distribution for Udio 130.}
  \label{fig:weight_udio130}
\end{figure}

\end{document}